\documentclass[english,aps,reprint,letter,amsmath,amssymb,superscriptaddress]{revtex4-1}
\usepackage{dcolumn}
\usepackage{bm}

\usepackage{hyperref,soul}% add hypertext capabilities
\usepackage{epstopdf}
\usepackage{color,bbm}

\usepackage[T1]{fontenc}
\usepackage{color}
\usepackage{graphicx}
\usepackage{babel}
\usepackage{ulem}

\begin{document}
\title{ Quantum simulation of indefinite causal order induced quantum
refrigeration}
\author{H. Cao$^{1,2}$}
\email{These two authors contributed equally to this work.}

\author{N. N. Wang$^{1,2}$}
\email{These two authors contributed equally to this work.}

\author{Z. A. Jia$^{1,2}$}
\author{C. Zhang$^{1,2}$}
\email{drzhang.chao@ustc.edu.cn}

\author{Y. Guo$^{1,2}$}
\author{B. H. Liu$^{1,2}$}
\author{Y. F. Huang$^{1,2}$}
\email{hyf@ustc.edu.cn}

\author{C. F. Li$^{1,2}$}
\email{cfli@ustc.edu.cn}

\author{G. C. Guo$^{1,2}$}
\address{$^{1}$CAS Key Laboratory of Quantum Information, University of Science
and Technology of China, Hefei, 230026, China ~\\
$^{2}$CAS Center For Excellence in Quantum Information and Quantum
Physics, Hefei, 230026, China}
\begin{abstract}
In the classical world, physical events always happen in a fixed causal order. However, it was recently revealed that quantum mechanics
allows events to occur with indefinite causal order (ICO).   In
this study, we use an optical quantum switch to experimentally investigate
the application of ICO in thermodynamic tasks. Specifically, we simulate the working system interacting with two identical thermal reservoirs in
an ICO, observing the quantum heat extraction even though
they are in thermal equilibrium where heat extraction is unaccessible by traditional thermal contact. Using such a process, we simulate
an ICO refrigeration cycle and investigate its properties. We also
show that by passing through the ICO channel multiple times, one can extract more heat per cycle and thus obtain a higher refrigeration performance. 
%Such a cooling machine only need to measure the control qubit on fixed basis, without resorting to manipulating the microscopic quantum state of the system. 
Our results suggest that the causal nonseparability can be a powerful resource for quantum thermodynamic tasks.
\end{abstract}
\maketitle
It is a deeply rooted concept that in a physical theory, there is
a well-defined pre-existing classical causal structure for which physical
events happen. However, from the Bell-Kochen-Specker
theorem \cite{bell1964einstein,kochen1975problem}, quantum
mechanics is incompatible with the viewpoint that  observables
have pre-existing values independent of the measurement. Inspired
by this, recent studies 
have shown that if we assume the causal relation to obey the laws
of quantum mechanics, it is possible for two events to occur with
superposed causal orders. Thus, there is no pre-existing causal relation
\cite{oreshkov2012quantum,brukner2014quantum}. The quantum causal
structure becomes especially crucial when quantum physics and general
relativity become relevant \cite{oriti2009approaches,hossenfelder2017experimental,marletto2017gravitationally,bose2017spin,peres2004quantum}.
A typical example is the quantum spacetime causal structure in the
study of quantum gravity \cite{christodoulou2019possibility,dewitt1967quantum,rovelli1990quantum,gambini2004relational,hardy2009quantum}.

Besides the fundamental properties of indefinite causal order
(ICO), the applications of ICO as an operational
resource in quantum protocols also attract considerable interests \cite{taddei2019quantum,jia2019causal}. It provides
remarkable enhancements ranging from channel discrimination \cite{chiribella2012perfect},
communication and computation complexity \cite{guerin2016exponential,feix2015quantum,araujo2014computational}
to quantum metrology \cite{zhao2020quantum,mukhopadhyay2018superposition}
, quantum information transmission \cite{ebler2018enhanced,salek2018quantum,chiribella2018indefinite}
etc. Recently, some of them have been experimentally studied by simulating ICO process with optical quantum switch \cite{procopio2015experimental,rubino2017experimental,goswami2018indefinite,guo2020experimental,goswami2020increasing,wei2019experimental}.

In thermodynamics, entropy in closed systems always tends to increase
definitely. An interesting question is what if applying
ICO in thermodynamic tasks. One of such example is the recent discovery
of ICO-based quantum refrigeration \cite{felce2020quantum}. There
are several different ways for refrigeration: the standard
one is powered by energy injected by a time-dependent driving force
\cite{campisi2015nonequilibrium,campisi2016dissipation}; the Maxwell
demon can steer the heat with feedback control
loops \cite{maruyama2009colloquium,elouard2017extracting}; while another
method is using invasive quantum measurements as a resource \cite{buffoni2019quantum}. All the above refrigeration
protocols works in pre-existing causal structures. The
ICO-based protocol provides a good supplement where no pre-existing
causal relation is assumed \cite{mukhopadhyay2018superposition,guha2020thermodynamic,felce2020quantum,markes2011entropy,simonov2020ergotropy,rubino2020time}.

In this paper, by faithfully adopting the protocol and extending the strategy in ref. \cite{felce2020quantum}, we experimentally simulate
the ICO induced heat extraction by optical quantum switch and investigate its feasibility to
construct a quantum refrigerator. We also show that by interacting
with reservoirs in ICO multiple times, one can extract more heat
from the reservoirs per cycle. The high accuracy achieved in our experiment
will motivate more operational protocols and  contribute
to broader research into ICO.

\textit{Protocol outline.}--- Consider a system with Hamiltonian
$\mathcal{H}$ and energy eigenstates $\left|n\right\rangle $ for
energy level $E_{n}$. After thermocontact with a thermal reservoir
with inverse temperature $\beta$, the resulting equilibrium state
of the system is always $T=e^{-\beta\mathcal{H}}/Z=\sum_{n}e^{-\beta E_{n}}/Z\left|n\right\rangle \left\langle n\right|$
regardless of the initial system state $\rho$, where $Z=\mathrm{Tr}\left(e^{-\beta\mathcal{H}}\right)=\sum_{n}e^{-\beta E_{n}}$
is the partition function. This thermodynamics operation can be characterized
by a completely positive trace preserving
(CPTP) map $\mathcal{T}\colon\mathcal{L}\left(\mathcal{H}\right)\rightarrow\mathcal{L}\left(\mathcal{H}\right)$
for which $\mathcal{T}\left(\rho\right)=T$ for all density operators
$\rho$. The Kraus decomposition is $\mathcal{T}\left(\rho\right)=\sum_{i}K_{i}\rho K_{i}^{\dagger}$,
where the Kraus operators $\left\{ K_{i}\right\} $ satisfy $\sum_{i}K_{i}^{\dagger}K_{i}=I$. 

Considering the situation where the system state $\rho$ undergoes
thermocontact sequentially with two identical thermal reservoirs. If we assume the definite causal order, then
the process is given either by $\mathcal{T}^{1}\circ\mathcal{T}^{2}\left(\rho\right)$
or $\mathcal{T}^{2}\circ\mathcal{T}^{1}\left(\rho\right)$, or potentially
a classical probabilistic mixture of them. The thermal state $T$ is always obtained. However, when applying ICO, the two events \textquoteleft thermocontact with thermal reservoir
1 firstly\textquoteright{} and \textquoteleft thermocontact with thermal
reservoir 2 firstly\textquoteright{} can occur in a superposed causal
order. An intriguing phenomenon arises---the resulting state
is different from $T$. Such an operation can be simulated
using the quantum switch \cite{chiribella2012perfect,chiribella2013quantum,procopio2015experimental}.
The action of ICO is achieved by routing particles through two channels
with the visiting order being tailored by the control qubit \cite{procopio2015experimental,rubino2017experimental,goswami2018indefinite,guo2020experimental,goswami2020increasing,wei2019experimental}.
When the control qubit $\left|\phi_{c}\right\rangle =\left|1\right\rangle $
($\left|0\right\rangle $) , the operations
$\mathcal{T}^{2}\circ\mathcal{T}^{1}$ ( $\mathcal{T}^{1}\circ\mathcal{T}^{2}$)
is carried out respectively. We denote the corresponding channel as
$\mathcal{S}^{T}$. In terms of Kraus operators, we have $\mathcal{S}^{T}\left(\rho_{c}\otimes\rho\right)=\sum_{ij}M_{ij}\left(\rho_{c}\otimes\rho\right)M_{ij}^{\dagger}$
and

\begin{equation}
M_{ij}=\left|0\right\rangle \left\langle 0\right|_{c}K_{i}^{1}K_{j}^{2}+\left|1\right\rangle \left\langle 1\right|_{c}K_{j}^{2}K_{i}^{1}\label{eq:(1)}
\end{equation}
where $K_{i}^{1}$ ($K_{j}^{2}$) represents the $i-$th ($j-$th)
Kraus operator of the thermalizing channels $\mathcal{T}^{1}$ ($\mathcal{T}^{2}$). 

Considering simplest non-trivial case, a two-level system,
the ground (excited) state is $\left|0\right\rangle $ ($\left|1\right\rangle $) with energy $E_{0}=0$ ($E_{1}=\Omega$),
thus, the Hamiltonian for the system is $\mathcal{H}=\Omega\left|1\right\rangle \left\langle 1\right|$.
The thermal state at a given temperature is $\rho=diag\left(1,e^{-\beta\Omega}\right)/Z$,
where $Z=1+e^{-\beta\Omega}$. In the following we set $\Omega=1$
for simplicity. If the ancillary control qubit is initialized as $\left|\phi_{c}\right\rangle =\left(\left|0\right\rangle +\left|1\right\rangle \right)_{c}/\sqrt{2}$.
the output state undergoing ICO with two identical thermalizing
channels is

\begin{eqnarray}
\mathcal{S}^{T}\left(\rho_{c}\otimes\rho\right) & = & \frac{1}{2}\left[\left(\left|0\right\rangle \left\langle 0\right|_{c}+\left|1\right\rangle \left\langle 1\right|_{c}\right)\otimes T\right.\label{eq:switch}\\
 &  & \left.+\left(\left|0\right\rangle \left\langle 1\right|_{c}+\left|1\right\rangle \left\langle 0\right|_{c}\right)\otimes T\rho T\right]\nonumber 
\end{eqnarray}
, where the control qubit is $\rho_{c}=\left|\phi_{c}\right\rangle \left\langle \phi_{c}\right|$.
 Note that control qubit gets entangled with system during ICO evolution; if the control
qubit is projected into $\left|\pm\right\rangle =\left(\left|0\right\rangle \pm\left|1\right\rangle \right)_{c}/\sqrt{2}$,
the system collapses into 
\begin{equation}
\mathrm{Tr}_{c}\left[\left|\pm\right\rangle \left\langle \pm\right|_{c}\mathcal{S}^{T}\left(\rho_{c}\otimes\rho\right)\right]=\frac{1}{2}\left(T\pm T\rho T\right)\label{eq:3}
\end{equation}
with probability $p_{\pm}=\frac{1}{2}\mathrm{Tr}\left[T\pm T\rho T\right]$.
Noting that the temperature of output system state could be different
from the thermal state $T$. This intriguing phenomenon suggests that the
ICO can either extract heat from or dump heat into the reservoir.

\begin{figure}
\includegraphics[scale=0.16]{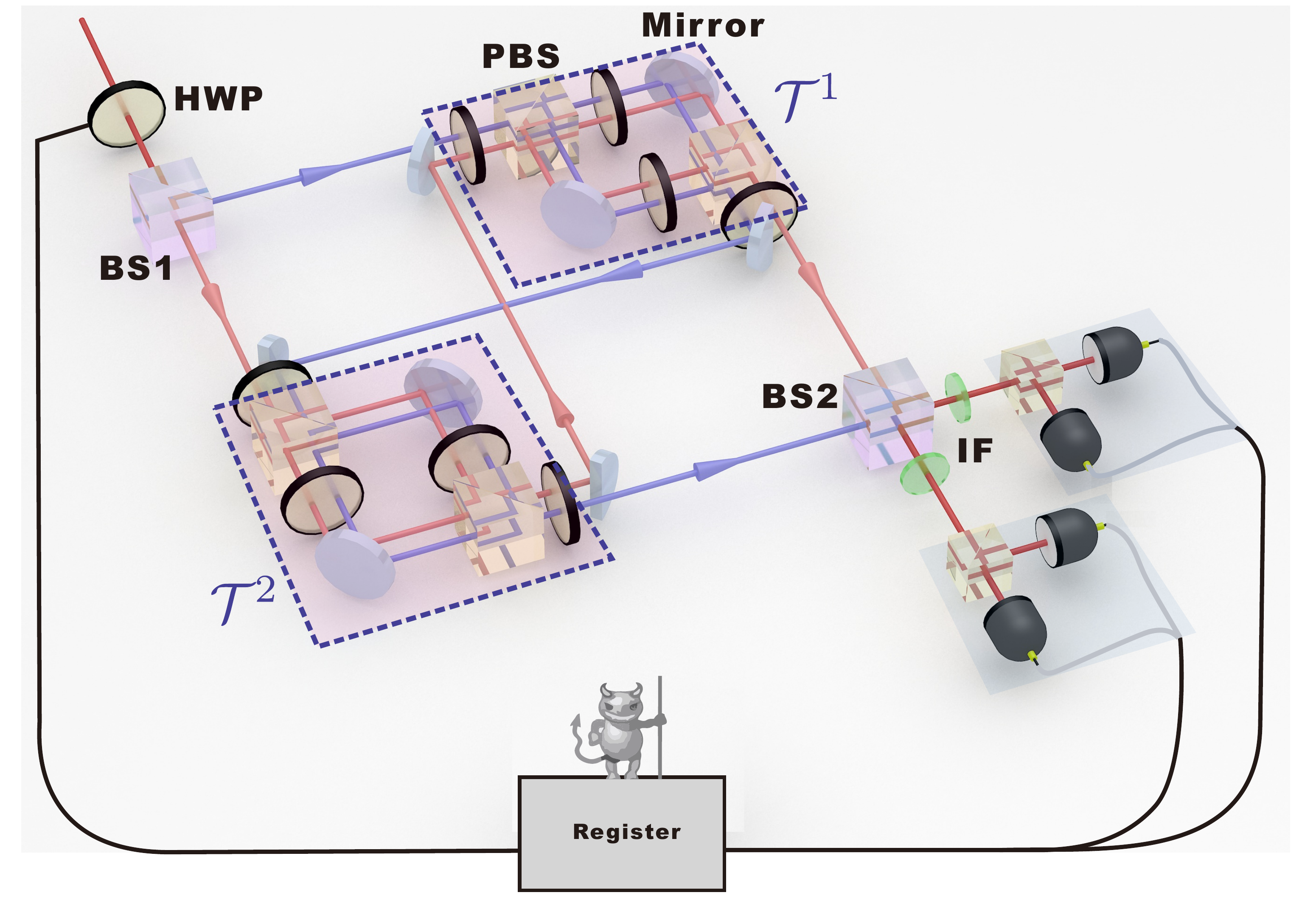}\caption{\label{fig:setup}Experimental apparatus. The quantum
switch contains two identical thermalizing channels (with pink planes
underneath) in a indefinite causal order. One of the causal order
is presented by a bluish optical path, the other is presented by
red one. BS: beam splitter,
PBS: polarization beam spliter, HWP: half wave plate, IF: interference filter.}
\end{figure}

\textit{Experimental implementation.}--- We simulate the ICO process with
tabletop photonic quantum switch. Twins photons at 780 nm are generated by spontaneous parametric down conversion and one of them is detected as trigger. The other one named heralded photon acts as working substance and then fed into a Mach-Zehnder interferometer(Fig \ref{fig:setup}).
We utilize photonic polarizations to mimic the two energy levels
of the working system, where horizontal (vertical) polarization state
H (V) represents the ground (excited) state  \cite{xu2014demon,mancino2017quantum}. In that way, The population of excited state represents the energy of the working system $\mathrm{Tr}\left(\rho\mathcal{H}\right)=e^{-\beta\Omega}/Z$ and hence reflects the temperature of the working system.
A system state at an arbitrary temperature is prepared by randomly
rotating the photonic polarization into H or V with a probability
proportional to its temperature. Energy detection can be realized
by measurements in Pauli $\sigma_{z}$ basis; thus, the temperature can be inferred. A beam splitter (BS1) introduces two spatial
modes as the control qubit. The polarization qubit undergoes the causal
order $\mathcal{T}^{1}\circ\mathcal{T}^{2}$ in one spatial mode,
while $\mathcal{T}^{2}\circ\mathcal{T}^{1}$ in the other. BS2
then coherently combines spatial modes and projects the control
qubit onto $\left|\pm\right\rangle $. It is reasonable to accept that the switch channel is accomplished when the two spatial mode form an interferometer \cite{wei2019experimental,guo2020experimental,procopio2015experimental}. While more rigorous verification of the ICO process requires measuring the witness \cite{rubino2017experimental}. A phase-locking
system is adopted to ensure the stability of the path interferometer,
with an average interferometric visibility of more than 99.7\% (See
Section II of Supplementary Material (SM) for details).  The interaction of system qubit with reservoirs can be modeled by two processes with the happening rate dictated by the temperature: (i) The qubit releases the energy to reservoir and decays to ground state; (ii) The qubit absorbs an excitation from reservoir and hops into the excited state. The generalized amplitude damping channel (GAD) explicitly simulates this interaction \cite{fisher2012optimal,lu2017experimental,mancino2017quantum}. The GAD is decomposed into $\ensuremath{\mathcal{T}\left(\rho\right)=\sum_{i=1}^{4}K_{i}\rho K_{i}^{\dagger}}$, of which the $\left\{ K_{1},K_{2}\right\}$  ($\left\{ K_{3},K_{4}\right\}$  ) forms a standard amplitude damping channel describing the process (i) ((ii)). For realization of sequential thermalizing channel, 16 Kraus operator settings $\left\{ K_{i}\otimes K_{j},i,j=1,2,3,4\right\}$  are randomly implemented and mixed classically, while two branches of interferometer is coherently combined without destroying the superposition We performed
the process tomography of the thermalizing channel at different temperatures. The average process fidelity exceeds 99.9\%, verifying
the credible simulation of the channel (section II of SM). 

\begin{figure}
\includegraphics[scale=0.073]{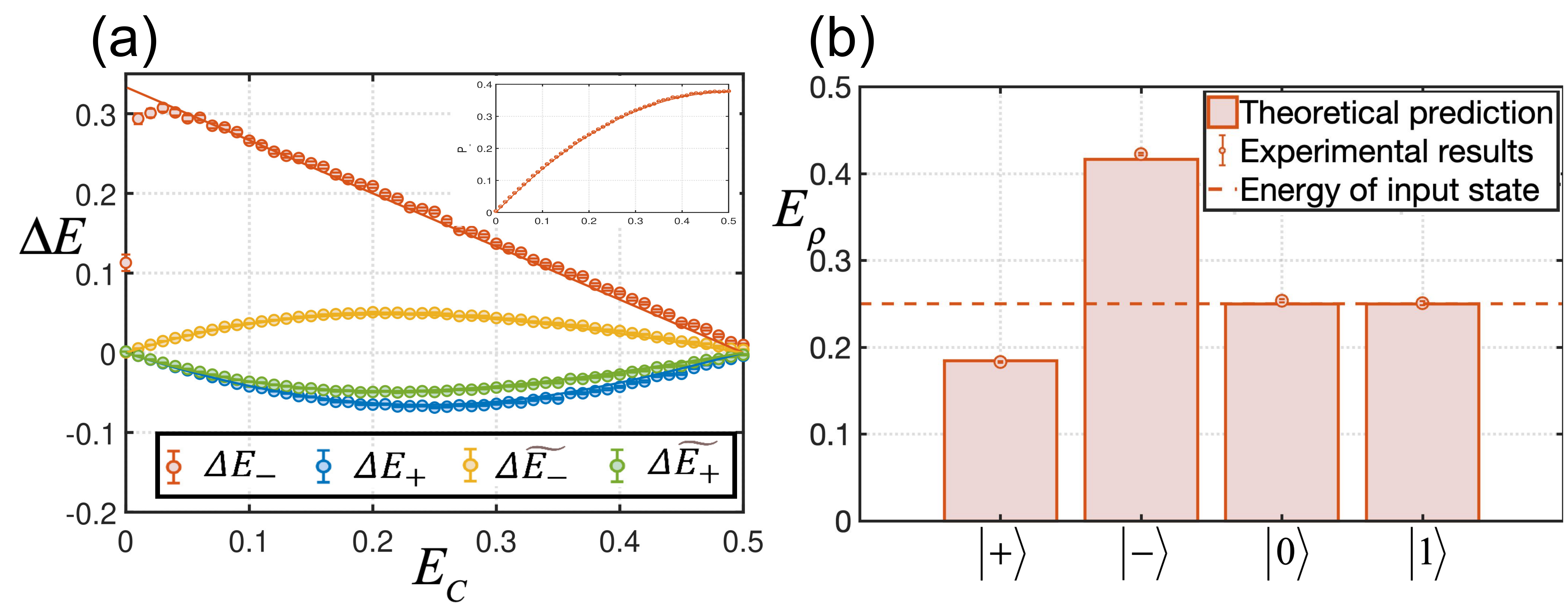}

\caption{\label{fig:result1}(a) Energy change $\Delta E_{\pm}=\mathrm{Tr}\left(\rho_{\pm}\mathcal{H}\right)-\mathrm{Tr}\left(\rho\mathcal{H}\right)$
where $\rho_{\pm}=\left[T\pm T\rho T\right]/\mathrm{Tr}\left[T\pm T\rho T\right]$
is the normalized output system state with the control qubit measured
in $\left|\pm\right\rangle $. The weighted energy change is presented
as $\Delta\widetilde{E}_{\pm}$. The horizontal coordinate
$E_{C}=e^{-\beta_{C}}/\left(1+e^{-\beta_{C}}\right)$ ranges from
0 to 0.5. (b) Energy of
the working system $E_{\rho}=\mathrm{Tr}\left(\rho\mathcal{H}\right)$
when the control qubit is measured at $\left|+\right\rangle ,\left|-\right\rangle ,\left|0\right\rangle ,\left|1\right\rangle $. The quoted error reflects the impact of Poissonian statistics
on the collecting count.}
\end{figure}

\textit{Results of the ICO process.}--- We first verify the non-classical
heat extraction driven by the ICO. We implement experiments
by traversing the temperatures of thermalizing channel. The system is
initialized into the thermal state with the same temperature ($1/\beta_{C}$)
as reservoir $\rho=T$. Fig. \ref{fig:result1} (a) shows the
measured energy change $\Delta E$ of the system qubit (as working
substance) after passing through the ICO channel. In the x-coordinate we use energy of the thermal state 
to represent the reservoir\textquoteright s temperature for simplicity.
By extrapolating experimental data ( red and blue dots in Fig.
\ref{fig:result1} (a)) to theoretical prediction (red and blue lines
in Fig. \ref{fig:result1} (a)), we find that the working substance
can extract the heat flow from the reservoir when the control qubit
is measured in $\left|-\right\rangle $ (initially prepared into $\left|+\right\rangle $), even though they initially
share the same temperature. This intriguing phenomenon is applicable
in arbitrary temperature cases except for zero and infinity.
The heat extraction decreases when the temperature increases, whereas
the successful probability $p_{-}=\mathrm{Tr}\left[\left|-\right\rangle \left\langle -\right|\mathcal{S}^{T}\left(\rho_{c}\otimes\rho\right)\right]$
increases (inset in Fig. \ref{fig:result1} (a)). The weighted energy
change $\Delta\widetilde{E}_{\pm}=p_{\pm}\Delta E_{\pm}$ is also
presented (orange and green dots and lines \ref{fig:result1} (a)). In the absence of information of control qubit's status by tracing it out, the averaging energy change strictly sums up to a vanishing value   $\Sigma_{k=\pm}\Delta\widetilde{E}_{k}=0$ both for experimental and theoretical data. This indicates that the heat extraction between thermal equilibrium systems can never occur spontaneously, in accord with Clausius statement of second law of thermodynamics \cite{pippard1964elements}.

For comparison, energy transfer with control
qubit measured in computational basis $\left\{ \left|0\right\rangle ,\left|1\right\rangle \right\} $ is also performed,
by exemplifying the reservoir\textquoteright s temperature $E_{C}=0.25$
(Fig.\ref{fig:result1}(b)). 
Such a case stands for the fixed causal order or equivalently the working substance classically contacting with reservoirs where they share a single temperature. This yield a trivial result that no heat extraction could achieve because classical thermocontact can not extract any heat flow from reservoirs at a single temperature however the limitation can be broken though by introducing ICO between reservoirs.

Obviously, the quantum heat extraction driven by ICO can be
used for thermodynamic tasks. For example, when working substance
appears at the heating component, we can interact it with an external
hot reservoir to release heat, thus refrigerating the
reservoir; Otherwise,
send it back to the reservoir to erase the unwanted heat exchange. An interesting question is whether
the working substance can become colder or hotter after passing through
the ICO channel multiple times.

\begin{figure}
\includegraphics[scale=0.066]{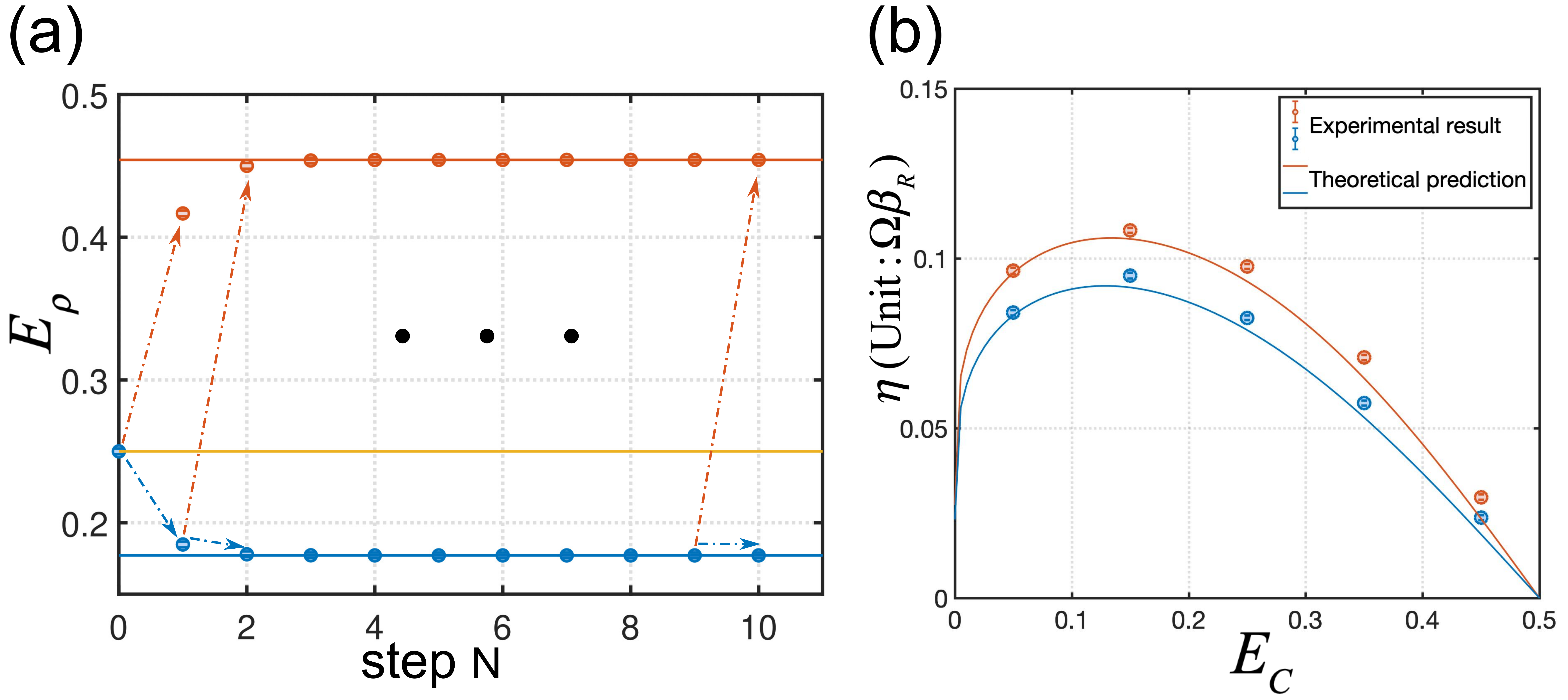}\caption{\label{fig:result2}(a) Energy of the working system after passing
through ICO channel multiple times. Each step will generate a
two-component outcome. The latter step is performed only when the former
step yields a failed outcome $\left|+\right\rangle $ (multi-pass
condition, as shown by the arrows in each step, with intermediate
ones are abbreviated with ellipsis). The saturated energy in both
outcomes is indicated by red and blue lines. (b) Coefficients of performance
based on the classical strategy (blue) and multi-pass strategy (red).
Here the coefficient is calculated in steady-state solution.}
\end{figure}

\textit{Results of multi-pass ICO.}--- Our second result is to investigate
this multi-pass strategy. We start with the working substance at the
same temperature as the reservoirs. As an example, we still adopt
the initial temperature to the one such that $E_{C}=0.25$. At each step, a single
run of quantum switch is carried out. Hence one initial state will
generate a two-component outcome (indicated by the arrows in Fig.\ref{fig:result2}(a)).\textcolor{blue}{{}
 }Here we only consider the case in which the working substance becomes
colder, and then send it into the next step as initial state. In the
experiment, since photons will be annihilated after being measured
in each step, we use the measurement results (classical information)
to determine the state preparation in the next step to simulate this
iterative process (as the loop depicted in Fig.\ref{fig:setup}).
The experimental results for the 10-steps ICO are summarized in Fig.\ref{fig:result2}(a),
in which the iteration process is indicated by the arrows. We observe
that when the multi-pass ($N\geq2$) ICO is implemented, a colder
working substance in the unwanted component could jump into a higher
temperature compared to in the single-step process ($N=1$). This
means that the multi-pass ICO may release the restriction for the
external reservoir for heat dumping. Interestingly, Fig.\ref{fig:result2}(a)
shows that the working substance will quickly saturate to a specific
temperature, which means the output working substance remains unchanged
in its input state when the control qubit is measured to be $\left|+\right\rangle $.
We theoretically calculate this steady-state solution for all temperatures
of the reservoir and also experimentally sample five points $E_{C}=\left\{ 0.05,0.15,0.25,0.35,0.45\right\} $,
finding that the working substance always tends toward this steady-state
solution after several iterations (section I of SM).

\begin{figure}
\includegraphics[scale=0.10]{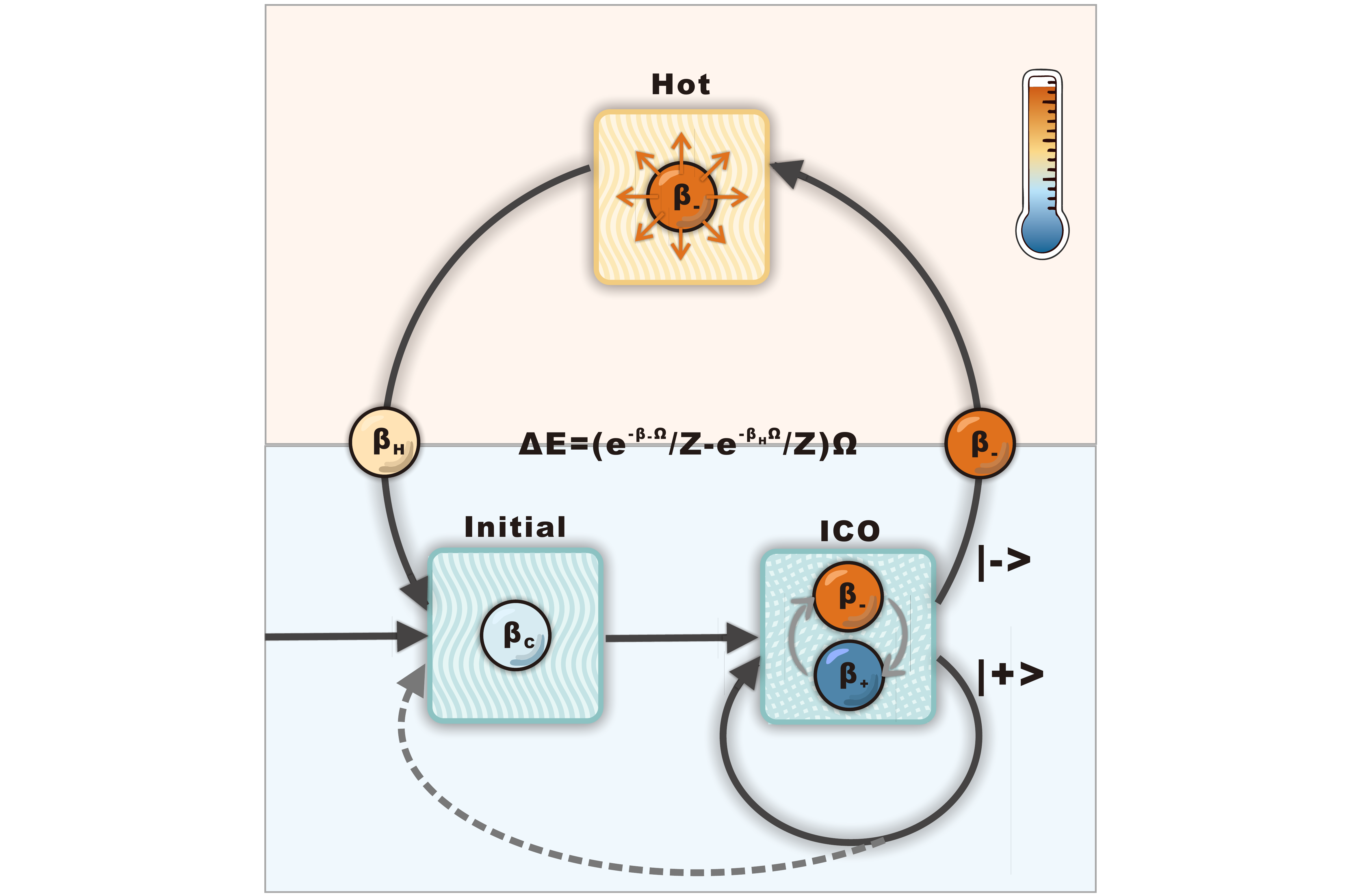}\caption{\label{fig:diagram}Diagram of single cycle of quantum
refrigerator. The reservoirs is denoted by squares, and the working substance is denoted
by the ball with the color corresponding to its temperature. Two recycling strategies are presented by the dotted 
arrow (classical strategy) and solid arrow (multi-pass strategy). The icon is taken from ref. \cite{rubino2020time}}
\end{figure}

\textit{Construction of a quantum refrigerator}.---  Considering the refrigeration task driven by ICO, an operational cycle is constructed with the diagram
of a single cycle shown in Fig.\ref{fig:diagram}. Stroke (i) Initialize the working substance by classically interacting
with the cold reservoir (preparing a colder working substance required
additional work cost, thus is excluded in our discussion). Then interacts
with the two cold reservoirs superposed in ICO. Stroke
(ii) Measure the control qubit. If the control qubit is collapsed
into $\left|-\right\rangle $, the working system successfully extracts
heat from the cold reservoir, followed by proceeding to the next stroke.
Otherwise, two
alternative strategies are available: (a) the working
system classically contacts the cold reservoir to recover its initialized
state, thereby undoing the unwanted heat change, and a new cycle is
implemented (termed classical strategy); (b) the ICO is repeatedly passed through until the desired outcome is obtained (termed multi-pass strategy). Stroke (iii) The working
substance makes classical thermocontact with the external hot reservoir
for heat release; subsequently, a new
cycle is started. 

To evaluate the performance of the quantum refrigerator, we introduce
the coefficient of performance, calculated by dividing the
heat change from the cold reservoir by the work cost of measurement
\cite{abdelkhalek2016fundamental}. The total heat extraction
per cycle $\Delta E$ is related to the temperature of particle entering
($\beta_{-}$) and leaving the external hot reservoir ($\beta_{H}$)
(depicted by the particles on the centreline). Basically the hot reservoir should not be hotter than the output working substance to release the heat so that it bounds the refrigerant range. Here, we fix  $\beta_{H}=\beta_{C}$ to maximize the heat
extraction, thus yields $\Delta E=E_{-}-E_{C}$. 

The magic of the ICO refrigerator can  be explained by Maxwell\textquoteright s
demon-like cooling mechanism. In each cycle of the quantum refrigerator,
the control qubit is measured with result stored in
register. A following operation is determined by the information
of register. However, The ICO scheme only relies on projection of control qubit on fixed basis and thermalizing channels alone, which releases the requirement of microscopic measurement and manipulation on working system (See Section VI in SM for detailed discussion). Since
we assume the control qubit has energy degeneracy, the measurement
itself does not cost energy. Rather, the energy cost comes from resetting
the register for proceeding the next
cycle, which refers to Landauer\textquoteright s erasure \cite{landauer1961irreversibility}.
The work cost is $\Delta W=\frac{1}{\beta_{R}}S$, where
$S=\left(p_{+}\ln p_{+}+p_{-}\ln p_{-}\right)$ is the Shannon entropy
of the register and $\beta_{R}$ is the inverse temperature of the
resetting reservoir. Therefore the coefficient of performance
is 
\begin{equation}
\eta=-\frac{\Delta E}{\bar{n}\Delta W}\label{eq:4}
\end{equation}
where $\bar{n}=\frac{1}{p_{-}}$ is the average number of measurements
consumed per cycle.

For classical strategy, the coefficient of quantum refrigerator
can be directly calculated by Eq. (\ref{eq:4}). For multi-pass strategy, it potentially contains many steps per cycle.
The heat flow and work cost in each step is not identical.
However, we can approximate the coefficient by assuming that the quantum refrigerator working at the
steady-state point, because a single cycle is much likely to undergo multiple steps and quickly evolve into equilibrium (It will lead to slight overestimation. For more rigorous comparison, realistic averaged coefficient is also provided in Section I of SM).  We comparatively present the coefficients 
under both strategies in Fig. \ref{fig:result2}(b),
where the solid lines (dots) show the theoretical prediction (experimental results) for sample temperatures $E_{C}=\left\{ 0.05,0.15,0.25,0.35,0.45\right\} $.
The underlying idea of two strategies is how to cancel the heat exchange in unwanted direction, either by classical thermocontact  to return or directly sending it into next ICO step. The results show that the multi-pass strategy surpasses
classical strategy, revealing the possibility of improving the efficiency of quantum thermodynamic tasks by introducing a more complex causal structure of the thermalizing channels.

\textit{Conclusions}.---  Our work
provides a new paradigm of quantum machine alternative to ones driven
by other non-classical features \cite{perarnau2015extracting,francica2017daemonic}. Despite of a measurement inside
strokes, both the outcomes are taken into account, hence the refrigerator
does not depend on post-selection to gain advantages. In addition, we provided a possible setup by adopting an equivalent circuit without projective measurement. where the feedback is controlled by control qubit (Section V in SM) \cite{felce2020quantum}.
This advantage of the ICO-driven protocol becomes crucial when the
control qubit cannot be used for direct thermalization, ICO process
allows us to access the free energy of control qubit.

The second law of thermodynamics imposes the irreversibility on thermodynamical evolution where as the system advances through
time, the heat can only naturally flow from hot to cold but not vice versa. Our experiment clearly shows
that, superposition of thermal operation in ICO can project the process onto diffrent direction of heat flow, which is something new for thermodynamics since previously the thermodynamics is established in pre-defined causal order. Noticing that the initial system state is Gibbs state and thermalizing channel is thermal operation of specific Hamiltonian \cite{chitambar2019quantum}. Both of them are resource free in thermodynamical resource theory. It shows interests to thermodynamical resource theory that development of thermodynamical resource by allowance of thermal operation superposed in ICO (as well as measure a control qubit) resulting in resourceful operation should be taken in to account. %\sout{more general framework without assuming the fixed classical causal structure can be developed. The quantum refrigerator only relies on projection of control qubit on fixed basis and feedback by thermalizing channels alone, in which the noninvasive measurement and manipulation on microscopic working system's state is exempt in contrast to traditional Maxwell demon} \cite{cottet2018maxwell,camati2016experimental,buffoni2019quantum} \sout{(Section IV of SM).} 
We expect that our work will advance further investigations
on the exotic property of ICO, as well as its
superiority in quantum thermodynamical tasks. Additionally, implementation of ICO based on high-performance optical quantum switch may encourage avenues on experimental research of other demanding ICO experiments like semi-device-independent certification of ICO \cite{bavaresco2019semi}, as well as measuring out-of-time-order correlation by adopting two superposed reversal sequential control order \cite{swingle2016measuring,zhu2016measurement}.  

\begin{acknowledgments}
We thank Yong-Xiang Zheng, Xue Li, Xiao Liu for beneficial discussions.
This work was supported by National Natural Science Foundation of
China (11734015, 62075208, 11774335, 11874345, 11821404, 11904357),
Anhui Initiative in Quantum Information Technologies (AHY070000, AHY020100,
AHY060300), National Key Research and Development Program of China
(2017YFA0304100, 2016YFA0301300, 2016YFA0301700), Key Research Program
of Frontier Sciences, CAS (QYZDY-SSW-SLH003), Science Foundation of
the CAS (ZDRW-XH-2019-1), the Fundamental Research Funds for the Central
Universities, Science and Technological Fund of Anhui Province for
Outstanding Youth (2008085J02).

\textit{Noted------ }Recently, we become aware of a related work
by Nie et al.\textcolor{blue}{{} }which experimentally study quantum
thermodynamics driven by ICO on nuclear spins using the nuclear magnetic
resonance system \cite{nie2020experimental}.
\end{acknowledgments}

%\normalem
%\bibliographystyle{apsrev}
%\bibliography{arxivICOQR}

\onecolumngrid 
\appendix

\section*{APPENDIX}

\section{Derivation of refrigerating performance}

%In this part we derive the refrigerating performance of passing the quantum switch multiple times.
{\it Steady-state solution.---} Here we derive the steady-state solution of the multi-pass ICO. Suppose the inverse temperature of the cold (hot) reservoir is $\beta_C$ ($\beta_H$), the input working qubit is in the state
\begin{equation}
\rho_{in} = \frac{1}{Z_x}\begin{pmatrix}
1 & 0 \\
0 & r_x
\end{pmatrix},
\quad
Z_x = 1+r_x,
\end{equation}
where $r_x=e^{-\beta_x}$. After interacting with two cold reservoirs in an ICO and projecting the control qubit into $|\pm\rangle$ basis, the resulting working qubit is
\begin{equation}
\rho_{\pm} = \frac{\rho_{\pm}'}{\textrm{Tr}[\rho_{\pm}']},\quad
\rho_{\pm}' = \frac{1}{2}\begin{pmatrix}
\frac{1}{1+r_C} \pm \frac{1}{(1+r_C)^2(1+r_x)} & 0 \\[2ex]
0 & \frac{r_C}{1+r_C} \pm \frac{r_C^2r_x}{(1+r_C)^2(1+r_x)}
\end{pmatrix},
\end{equation}
with probability
\begin{equation}
p_{\pm}=\textrm{Tr}[\rho_{\pm}']=\frac{1}{2}\left[1 \pm \frac{1+r_C^2r_x}{(1+r_C)^2(1+r_x)}\right].
\end{equation}

\begin{figure}[tb]
\centering
\includegraphics[width=0.5\textwidth]{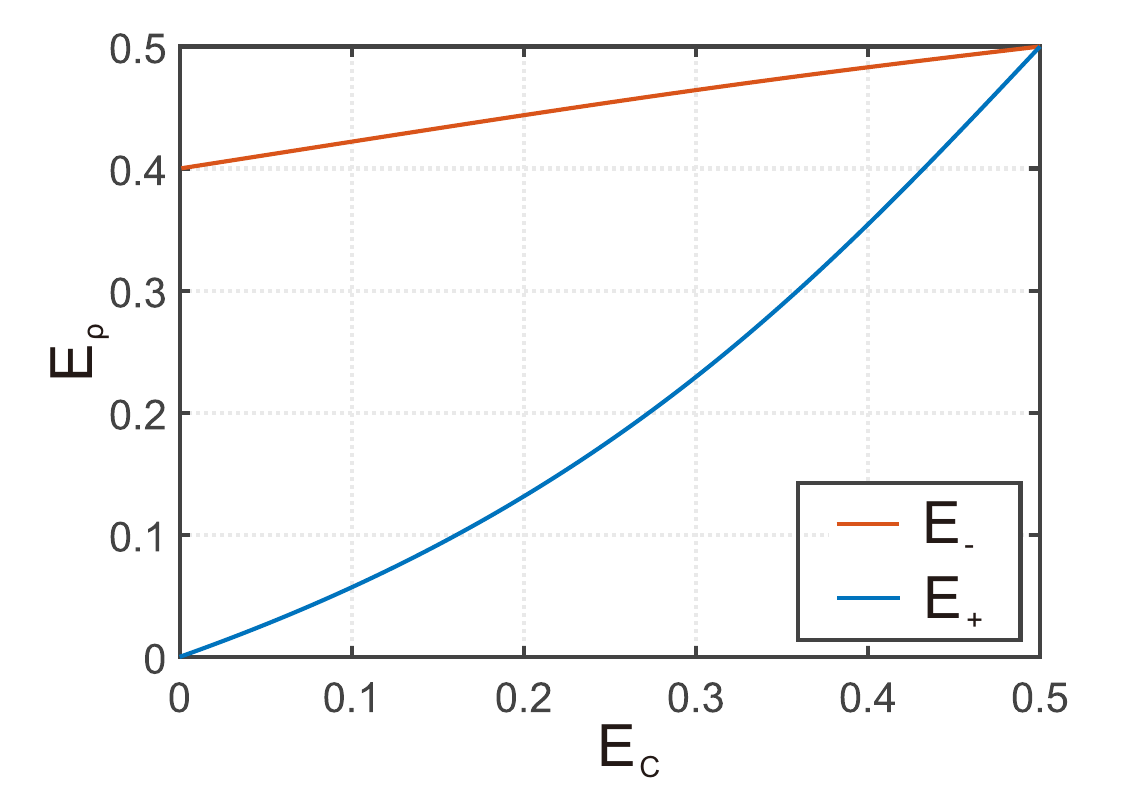}
\caption{\label{Fig:S1} Red and blue lines represent the steady-state temperatures of the working system when the control qubit is measured to be $|\pm\rangle$ respectively. }  \end{figure}

In the refrigeration cycle, the working system is initially in a temperature equal to the cold reservoir. Then the working system is passed through the ICO channel many times until the control qubit is measured in $|-\rangle$. If the control qubit is measured to be $|+\rangle$, we send back the working system to the ICO channel. We iterate this process several times and find that it quickly approach a steady-state solution, which means the working system remains unchanged when the ICO process is failed
\begin{eqnarray}
\rho_+ &=&\rho_{in}\\\nonumber
\Longrightarrow r_x&=&\sqrt{(1-r_C)^2+r_C}-(1-r_C).
\end{eqnarray}
At this point, the temperature (energy) of the output working system is
\begin{eqnarray}
E_+=e^{-\beta_+}/Z_+&=&\frac{r_x}{1+r_x}\\\nonumber
E_-=e^{-\beta_-}/Z_-&=&\frac{3r_x^3+11r_x^2+11r_x+2}{5r_x^3+22r_x^2+22r_x+5}.
\end{eqnarray}
We plot this two steady-state temperatures as a function of the cold reservoir's temperature as shown in Fig.\ref{Fig:S1}.

\begin{figure}[tb]
\centering
\includegraphics[width=0.5\textwidth]{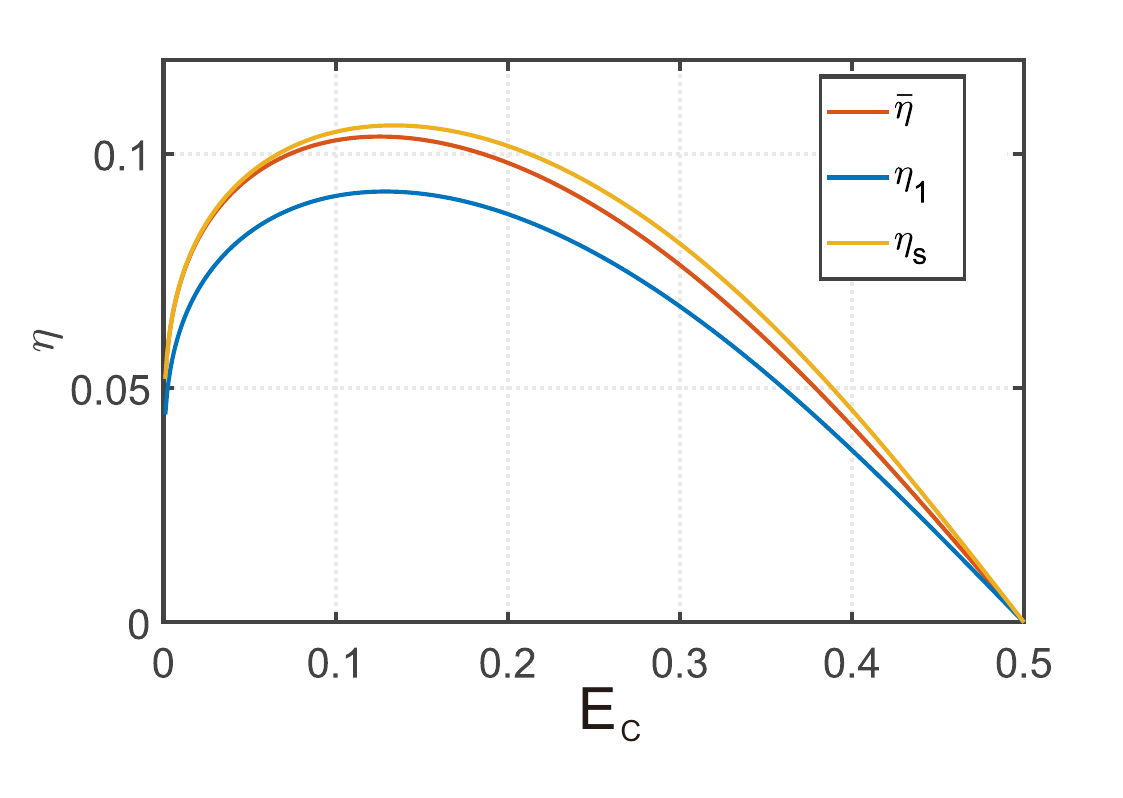}
\caption{\label{Fig:S2} The refrigerating coefficient with different temperature of the cold reservoir.  $\bar{\eta}$ is the averaged coefficient of the multi-pass ICO refrigerator. $\eta_{s}$ is the coefficient when the refrigerator is working at the steady-state point. $\eta_1$ is the coefficient for the first cycle, which means when the process failed, we interact the working system with the cold reservoir classically to reset its temperature to the initial value.   }  \end{figure}

{\it Refrigerating performance.---} The performance of the multi-pass ICO refrigerator can be calculated by dividing the averaged heat transfer by the averaged work cost. In each cycle, the work cost comes from erasing the control qubit's information in order to return to its initial state. If the measurement result of the control qubit is recorded in a register with inverse temperature $\beta_R$, the work cost is
\begin{equation}
\Delta W^n=\frac{1}{\beta_R}(p_+^n\textrm{ln}p_+^n+p_-^n\textrm{ln}p_-^n),
\end{equation}
here the superscript n denotes the nth cycle. The net heat transfer in each cycle from the cold bath to the hot bath is
\begin{equation}
\Delta E^n=E^n_--E_H,
\end{equation}
then the averaged refrigeration coefficient can be written as
\begin{equation}
\bar{\eta}=-\frac{\sum_{i=1}^\infty P^i \Delta E^i}{ \sum_{i=1}^\infty P^i \sum_{k=1}^i \Delta W^k},
\end{equation}
where $P^i=\prod_{j=1}^{i-1} p_+^j \cdot p_-^i$ is the probability that the working system passes through the ICO channel i times. In fact, this averaged coefficient is very close to the coefficient that the refrigerator is working at the steady-state point (the subscript s means steady)
\begin{equation}
\eta_{s}=-\frac{\Delta E^{s}}{\Delta W^{s}/p_-^{s}}.
\end{equation}
where $1/p_-^{s}$ denotes the averaged number of times that the working system passes through the ICO channel. We compare these efficiencies in Fig.\ref{Fig:S2}.

%%%%%%%%%%%% FIGURE S1 %%%%%%%%%%%%%%%%%%%%%%%%%%%%%%%%%%
%%%%%%%%%%%%%%%%%%%%%%%%%%%%%%%%%%%%%%%%%%%%%%%%%%%%%%%%

%%%%%%%%%%%% FIGURE S2 %%%%%%%%%%%%%%%%%%%%%%%%%%%%%%%%%%
%%%%%%%%%%%%%%%%%%%%%%%%%%%%%%%%%%%%%%%%%%%%%%%%%%%%%%%%

\section{Linear optical quantum simulator}

{\it Quantum simulation protocol.---} Quantum simulation is put forward to overcome the difficulty in exploring some less controllable or accessible quantum system by using other controllable ones. In quantum simulation protocol, the targeted physical system to be
surveyed is mapped into the simulating system, and the desired evolution can be mapped into the quantum evolution implemented on the simulating system under specific physical conditions. While the targeted physical feature is investigated by observing the simulating system. In our work, the targeted physical system is the two-level qubit embedded in the ICO process. The desired evolution is the two identical thermalizing channel with superposed causal order. The physical condition is that the thermal channel is isochoric and its corresponding thermal reservoir is sufficient large. That is, the change in energy of working system stems from the heat flow from the reservoirs, while the heat exchange in single cycle by a microscopic working substance does not affect the temperature of large thermal reservoir. The simulating system is photonic qubit in linear optical system, where the temperature (or energy) is evaluated by the population on the qubit eigenstates. The desired evolution can be effectively simulated by decomposing the thermalizing channel into generalized amplitude damping channel, and adopting the optical quantum switch to emulate the superposed quantum spacetime. The ICO quantum simulator greatly benefits from the experimental advantages of linear optical system. Although there are a variety of seminal results regarding quantum thermodynamics in terms of trapped ion or atom \cite{rossnagel2016single,hu2020quantum,raizen2009comprehensive},  superconducting circuit \cite{quan2006maxwell,cottet2017observing}, field-effect transistors \cite{chida2017power}, Nuclear Magnetic Resonance (NMR) spectroscopy \cite{camati2016experimental}, a highly controlled interaction with surroundings is still quite a challenge as residual noise of surrounding is inevitable.  Photonic qubit is highly resistant to external noise. This enables us to accurately simulate desired evolution without perturbed by unwanted noise. Besides, a real indefinite causal order corresponding to superposed quantum spacetime is expected to arisen when both general relativity and quantum physics become relevant \cite{brukner2014quantum,marletto2017gravitationally}, however so far no direct experimental evidence has been observed. A closest alternative approach is adopting quantum switch to control the order in which the channel is applied. Thus in our work, the thermal channels embedded in optical switch has been an optimal choice enabling us to accurately and faithfully simulate the ICO-based thermaldynamical task.

{\it Thermal state preparation.---} Conceptually, the temperature of a ensembles is directly related to the internal energy, while the energy of a two-level quantum system account for the population ratio on its energy levels. Concretely, in our work, For a presumed system Hamiltonian $\mathcal{H}=\varOmega\left|1\right\rangle \left\langle 1\right|$, the thermal state at inverse temperature $\beta$,  $\rho=e^{-\beta\mathcal{H}/Z}=diag\left\{ 1,e^{-\beta\mathcal{\varOmega}}\right\} / \left( 1+e^{-\beta\varOmega}\right)$, bridges its population, temperature and energy. The two energy level of quantum particle is simulated by horizontal and vertical
polarization. Although the polarizations are degenerate, the 'energy' of simulating qubit can also be evaluated by its populations, thus it is capable of simulating the thermal state of this Hamiltonian, with the temperature or energy change during thermodynamical process being reflected by the population change. 

To experimentally prepare the initial working system's thermal state $\rho=T=\left[\begin{array}{cc}
1 & 0\\
0 & e^{-\beta\varOmega}
\end{array}\right]/\left(1+e^{-\beta\varOmega}\right)$, A half wave plate (HWP) mounted in motorized rotation mount interchangeably implements two orthogonal unitary operations which is driven by a sequence of random logic number of which the probability is proportional to the ground (excited) population $\frac{1}{1+e^{-\beta\varOmega}}$ ($\frac{e^{-\beta\varOmega}}{1+e^{-\beta\varOmega}}$). The two unitary operations respectively set the single photon into horizontal and vertical polarization. The target thermal state is achieved by averaging the results during the data analysis \cite{ringbauer2018certification}.

{\it Thermalizing channel.---} The interaction of a qubit system with a thermal bath definitely compel arbitrary initial system state into a certain final state , which can be emulated by a generalized amplitude damping (GAD) channel. The corresponding map is composed of four Kraus operators:
\begin{equation}
K_1 = \sqrt{p}\begin{pmatrix}1 & 0 \\ 0 & \sqrt{\gamma} \end{pmatrix},K_2 = \sqrt{p}\begin{pmatrix}0 & \sqrt{1-\gamma} \\ 0 & 0 \end{pmatrix},K_3 = \sqrt{1-p}\begin{pmatrix}\sqrt{\gamma} & 0 \\ 0 & 1 \end{pmatrix},K_4 = \sqrt{1-p}\begin{pmatrix}0 & 0 \\ \sqrt{1-\gamma} & 0 \end{pmatrix}
\end{equation}
the first two operators describe the process that the qubit decays to the ground state transferring its energy to the reservoir (which belongs to a standard amplitude damping channel), while the latter two describe the opposite process which the qubit absorbs energy from the reservoir and hops to the excited state. The probability of the two processes is determined by the temperature of the reservoir and the energy gap of the two-level system. The other parameter $\gamma$ represents the damping rate of the two processes which is determined by the interaction time between qubit system and reservoir. Here we only interested in the infinite interaction time, so $\gamma$ is 0.

\begin{figure}[ht]
\includegraphics[width=0.9\textwidth]{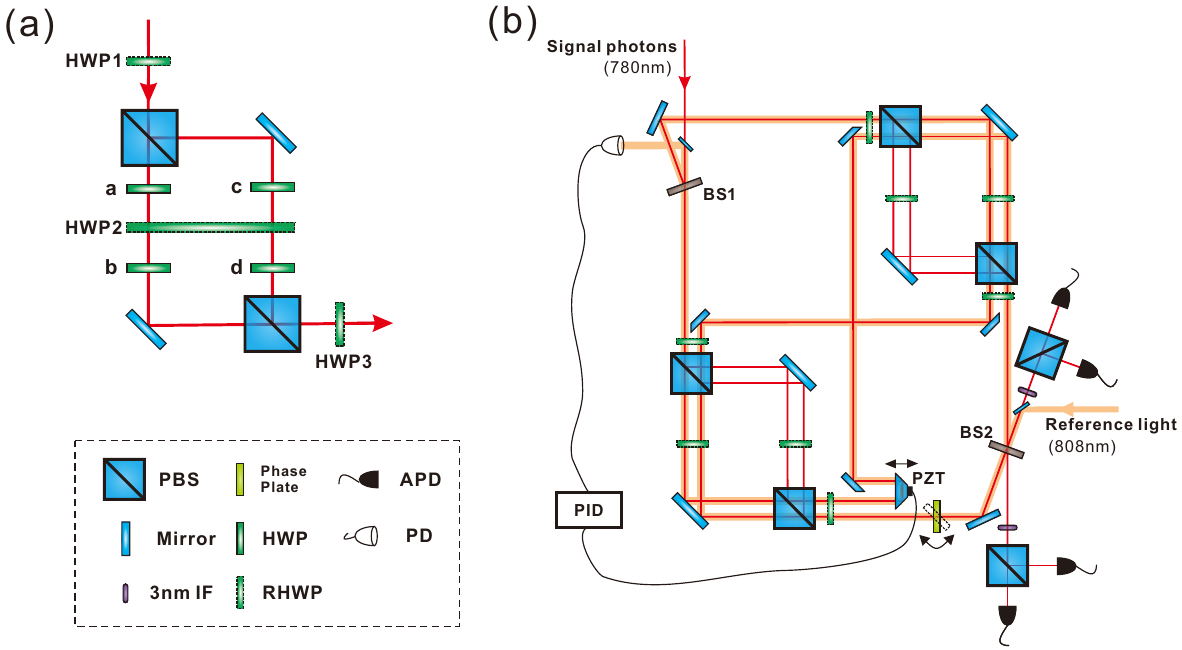}
\caption{ \label{fig:locking} (a) The detailed experimental setup of the GAD channel. (b) Our phase-locking system, which consists of an auxiliary power-stabilized diode laser, a photon detector and a piezo-transducer (PZT) controlled by a PID regulator. The BS1 and BS2 form the block MZ interferometer. Here we use $0^\circ$ BS in order to achieve the same reflection transmission ratio for both polarizations. The reference light (thick orange line) and the signal light (thin red line) travel in opposite directions and have different heights in space. The reference light doesn't pass through the HWPs in the GAD channel. Symbols used in the figure are PBS, polarization beam splitter; IF, interference filter; HWP, half wave plate; RHWP, rotatable HWP; APD, avalanche photodiode; PD, photon detector.}
\end{figure}

The experimental setup to realize the GAD channel is depicted in Fig.\ref{fig:locking} (a), which consists of a Mach-Zenhder (MZ) interferometer and several randomly rotated waveplates. The $|H\rangle$ ($|V\rangle$) polarization state of the photon is served as the ground (excited) state of the working qubit. The angles of the four fixed HWPs are set to $\cos 4a=\cos 4b=\sqrt{r}$, $c=0$ and $d=\frac{\pi}{2}$. One can easily check that the angles of HWP1-3 to realize the four Kraus operators in Eq.B1 are ($\frac{\pi}{4},0,\frac{\pi}{4}$), ($\frac{\pi}{4},\frac{\pi}{4},0$), ($0,0,0$), and ($0,\frac{\pi}{4},\frac{\pi}{4}$), respectively. The three HWPs are mounted on the motorized rotation stages and controlled by a computer, which allows us to randomly switch among the angle settings to realize each Kraus operator with desired probability $\overrightarrow{p}=\left\{ 
\sqrt{p},\sqrt{p},\sqrt{1-p},\sqrt{1-p}\right\} $. A generalized amplitude damping channel can be engineered by proper mixture of these four Kraus operator and tracing over the classical information about which Kraus operator is implemented.

\begin{figure}[ht]
\centering
\includegraphics[width=1\textwidth]{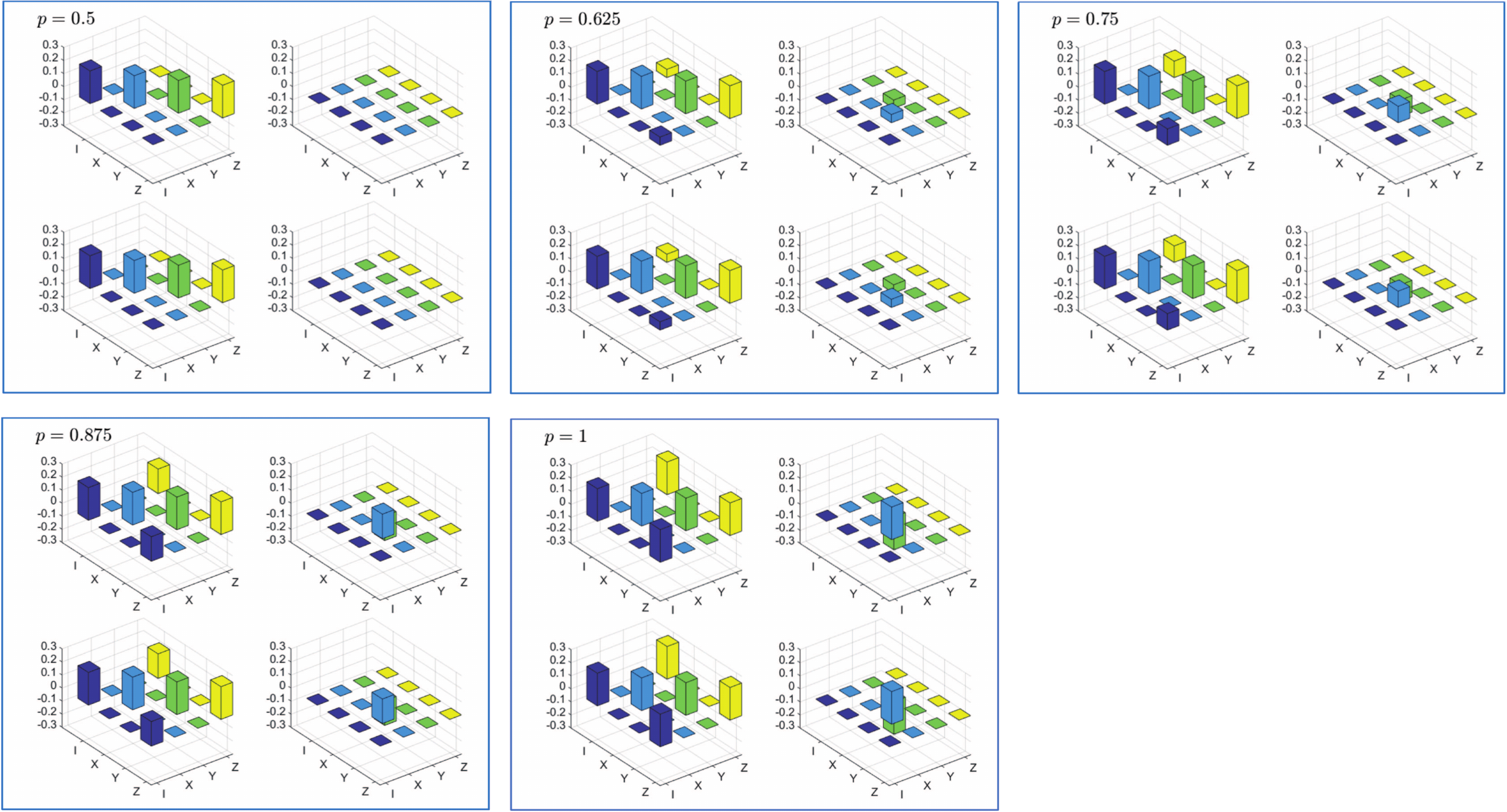}
\caption{\label{fig:S3} Results of single qubit process tomography in pauli basis. In each box, the top two are real(left) and imaginary(right) part of reconstrcted process matrix, and the bottom is the theoretical result.}
\end{figure}

To characterize the channel, we perform single qubit process tomography with different transition probabilities $p=\{0.5, 0.625, 0.75, 0.875, 1\}$. The reconstructed process matrices are shown in Fig.\ref{fig:S3} . The process fidelities are calculated to be $0.99976 \pm 0.00005, 0.99986 \pm 0.00003, 0.99968 \pm 0.00004, 0.99937 \pm 0.00005$ and $0.99832 \pm 0.00007$ respectively, testifying the credible simulation of the channel.

\begin{figure}[ht]
\includegraphics[scale=0.5]{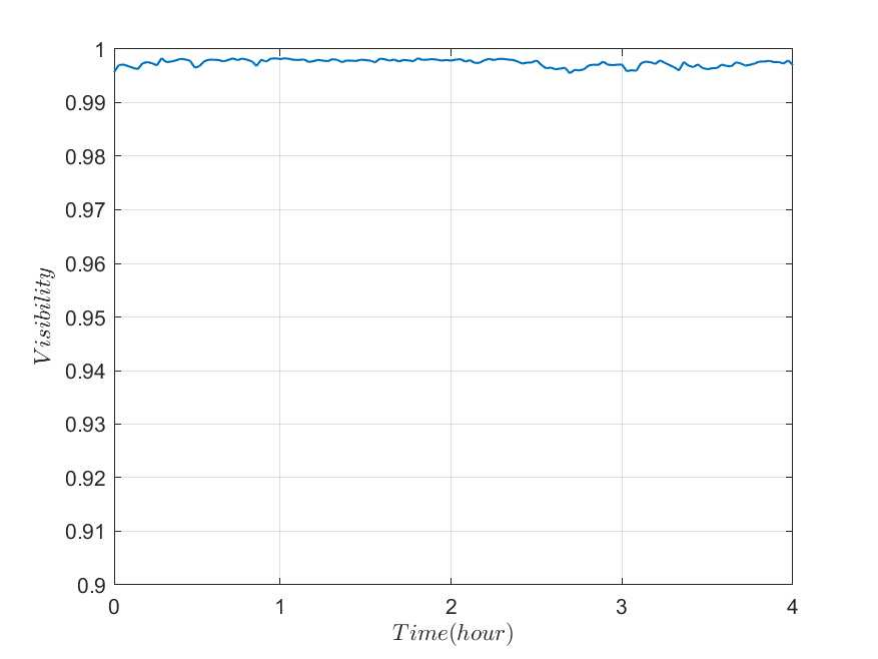}
\caption{\label{fig:S4} Result of interference visibility of the signal photons for four hours.}
\end{figure}

{\it Phase locking.---}Our optical quantum switch is based on a 1 meter long block MZ interferometer. An active phase-locking system is employed to compensate the phase noise in the interferometer. To reduce the impact on the signal photons, the reference light at 808 nm is traveled in an opposite direction, the detailed experimental setup is shown in Fig.\ref{fig:locking}(b). After phase-locking, we measure the interference visibility of the signal photons at the output ports of BS2 for four hours and observe that a high visibility of 0.997 is maintained during the experiment (Fig.\ref{fig:S4}).

\begin{figure}[ht]
\centering
\includegraphics[width=1\textwidth]{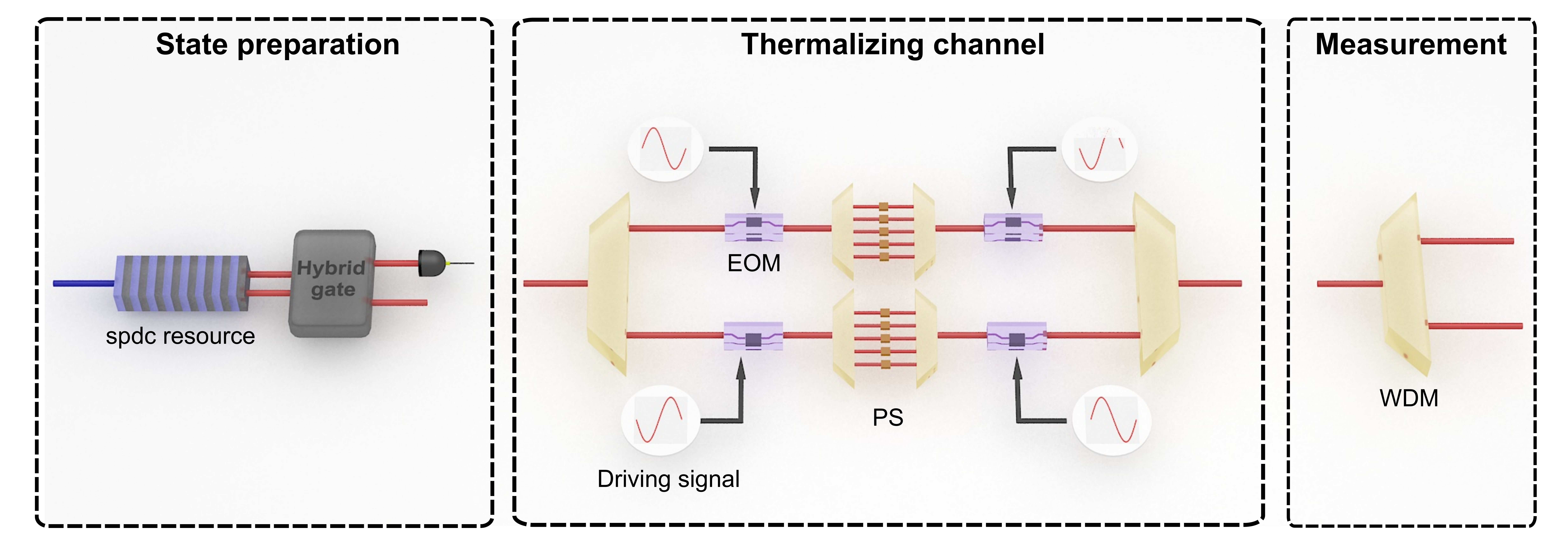}
\caption{\label{fig:frequency}schematic of optical simulator of ICO thermodynamics in terms of frequency encoding. Three fundamental ingredients are respectively drawn in black dotted boxes.}
\label{}
\end{figure}

{\it Frequency-bin qubit encoding.---} In the optical simulator of indefinite causal order driven thermodynamics in main text, the energy level is encoded on the eigenstates of photonic polarization, which is energy degenerate. Although the physical nature can be faithfully emulated, degenerate energy level prohibits a real energy exchange during thermalizing process, restricting its further application. Here we provide an alternative way by employing frequency-bin encoding where the discrete frequency-bin is non-degenerate and closely relates to energy of photon. The theoretical model can still be justified in frequency-bin coding, while the setup for manipulating working system states should be transferred into corresponding basic elements.

(i) State preparation. For any feasible encoding manner, a universally tailored single photon or multiphoton preparation is particularly demanding. It was demonstrated that a genuine discrete color-entangled state $\left|\psi\right\rangle =\alpha\left|\omega_{1}\right\rangle _{s}\left|\omega_{2}\right\rangle _{i}+\beta\left|\omega_{2}\right\rangle _{s}\left|\omega_{1}\right\rangle _{i}$ with tunable population ratio $\alpha,\beta$ is accessible by Sagnac-loop SPDC resource combined with hybrid quantum gate \cite{ramelow2009discrete}, or spontaneous four-wave mixing inside micro-ring cavity \cite{reimer2016generation,kues2017chip}, where the $\omega_{1},\omega_{2}$ denotes the two discrete frequency-bin. An artificial thermal state $\rho=diag\left\{ \alpha,\beta\right\} $ can be prepared by detecting one of the twins photons as trigger and averaging over its frequency information. The brief sketch is shown in Fig.\ref{fig:frequency}.

(ii) Thermalizing channel construction. In frequency encoding optical simulator, the thermalizing channel can also be simulated by randomly switching among its Kraus operators with corresponding probability. 
Experimental endeavor dictates that EOM and pulse shaper (PS) enable us to implement universal single photon gate on frequency-bin qubit by EOM-PS-EOM architecture \cite{lu2020fully}. To construct the thermalizing channel, a wavelength-division multiplexing (WDM) first separates the two frequency-bins. one EOM-PS-EOM setup is inserted in each wavelenngth channel to individually modulates two frequency-bins. The EOM-PS-EOM setup is randomly switched between the identity $I$ and Pauli-x $\sigma_x$ operator with desired probability corresponds to the temperature of reservoir. The identity operator $I$ leaves the frequency bin unchanged while the $sigma_x$ make the frequency bin hopping into another one. Because the EOM and phase PS are both driven by the electric signal, it is feasible to fast switch between these operators.

(iii)Measurement. In our protocol, only the measurement on energy basis $\left\{\left|\omega_{1}\right\rangle ,\left|\omega_{2}\right\rangle \right\}$ is performed, which can be simply realized by a WDM followed by single photon detectors.

\section{Exploration on possibility of quantum heat extraction with different settings}

The indefinite causal order has recently been proposed as a novel resource for wide-ranging applications with regard to quantum communication \cite{chiribella2012perfect,salek2018quantum,guerin2016exponential} and quantum computation \cite{araujo2014computational}. However, to what extent these advantages are specific to indefinite causal order, rather than generic to coherently superposed quantum operations, has been a subject of recent debates \cite{salek2018quantum,abbott2020communication,guerin2019communication,chiribella2019quantum}. It has been pointed out that coherent superposition of noisy channels could also reduce the noise in quantum communication \cite{abbott2020communication,guerin2019communication}. This excludes the necessity of indefinite causal order in evading the noise in communication tasks.  Exploring weather or not such a proposition is still hold in other contexts, especially in quantum dynamical tasks, is particularly demanding.

\begin{figure}[ht]
\includegraphics[scale=0.12]{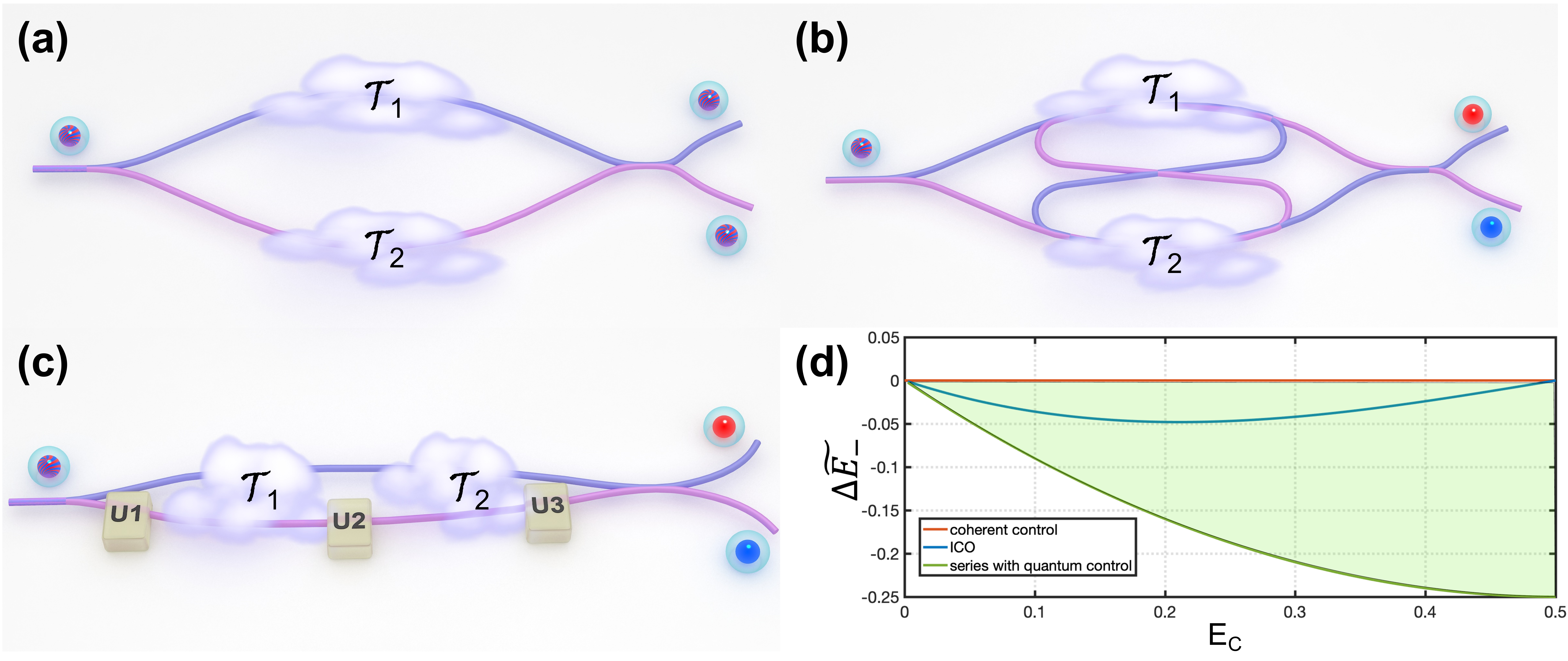}
\caption{\label{fig:comparisonsetting}Different schemes for realizing heat extraction out of specific thermal reservoir. (a) Coherent control of channels. (b) Thermalizing channel superposed in indefinite causal order. (c) Channels in series with quantum control. In all schemes, superposition of trajectories is necessary. In (c), universal research of unitaries is performed and the optimal case occurs when $U_1=U_2=\mathbb{I}$ and $U_3=\left(\begin{array}{cc}
1 & 0\\
0 & e^{i\pi}
\end{array}\right)$. In such case, the ability of heat extraction for above schemes is subject to the order $(c)>(b)>(a)$. (d) Comparison of ability on heat extraction for three schemes. In each scheme, various thermal state for initial working system and thermal reservoir is studied. The initial working system state is  represented by its energy $E_c$ in horizontal coordinate while the heat extraction is quantified by energy change of system state $\Delta E$ in vertical coordinate conditioned at control qubit projected into desired state.}
\end{figure}

In this section, we rigorously compare the ability of different schemes to realize heat extraction out of specific thermal reservoir. We reiterate the three layouts (Fig.\ref{fig:comparisonsetting}) which have been previously surveyed for quantum communication tasks \cite{rubino2021experimental}, a) coherent control of channels, b) channels superposed in indefinite causal order, c) channels in series with quantum controlled operations.

{\it Coherent control of channels.---}In the first layout, we revisit coherent control of quantum channels by employing two independent identical thermal channel placed in parallel. A control system determines which channel is used to interact with working system. Such approach was originally raised for error filtration and served as an approach to control unknown unitaries \cite{gisin2005error,araujo2014quantum}. The visual sketch is shown in Fig.\ref{fig:comparisonsetting}(a). Similar to the ICO scheme (Fig.\ref{fig:comparisonsetting}(b)), two independent identical thermalizing channels and quantum switch are used in coherent control of channel scheme. While instead of sequentially passing through both channels with order routed by control system, the working system only undergoes either channel determined by control system. A complete description of the coherent control of channel $\mathcal{S}^{T}$is formulated as
\begin{eqnarray}
\mathcal{S}^{T}\left(\rho_{c}\otimes\rho\right)&=&\sum_{ij}M_{ij}\left(\rho_{c}\otimes\rho\right)M_{ij}^{\dagger}\\\nonumber
M_{ij}&=&\left|0\right\rangle \left\langle 0\right|_{c}K_{i}^{1}+\left|1\right\rangle \left\langle 1\right|_{c}K_{j}^{2}
\end{eqnarray}
where control system $\rho_{c}$ and working system $\rho$ is initialized into a product joint state, and ${K_{i}}({K_{j}})$ completely forms the $i(j)$ th thermalizing channel. The notions here are the same as eq.(1) in main text. By summing over each choice of $\left\{ K_{i},K_{j}\right\}$, the control-system joint state evolves into
\begin{eqnarray}
\mathcal{S}^{T}\left(\rho_{c}\otimes\rho\right)=\frac{\mathbb{I}}{2}\otimes T+\frac{1}{8}\left[\left|0\right\rangle \left\langle 1\right|+\left|1\right\rangle \left\langle 0\right|\right]\mathcal{O}\rho\mathcal{O}^{\dagger}
\end{eqnarray}
Here $\mathbb{I}$ is identity operator for control qubit and $T=\sum_{i}K_{i}\rho K_{i}^{\dagger}=\left(\begin{array}{cc}
1 & 0\\
0 & e^{-\beta\Omega}
\end{array}\right)/ \ \left( 1+e^{-\beta\Omega} \right)$ is the thermal state corresponding the thermalizing channel, and $\mathcal{O}=\sum_{i}K_{i}$. The normalized postmeasured system state conditioned at $\left|\pm\right\rangle$ for control qubit is given by 
\begin{eqnarray}
\rho_{\pm}=\left[\frac{T}{2} \pm \frac{ \mathcal{O} \rho \mathcal{O}^{\dagger}}{8} \right]/ \mathrm{Tr}\left[\frac{T}{2} \pm \frac{ \mathcal{O} \rho \mathcal{O}^{\dagger}}{8}\right]
\end{eqnarray}
It is clear that the output state partially depends on the initial state, but here the aim of our scheme is to investigate the internal energy change of system state rather than how much quantum information survive the noise. It is interesting to discuss the energy of system state $\mathrm{Tr}\left(\rho_{\pm}H\right)$ after the action of coherent control of channels.  Considering the initial system state prepared as thermal state $T$, we numerically calculate the weighted energy change $\triangle\widetilde{E}_{-}=p_{-} \left[ \mathrm{Tr}\left(\rho_{-}H\right)-\mathrm{Tr}\left(\rho H\right) \right]$ undergoing coherent control of channels. We traverse the temperature of initial state $T$ with the energy ranging from 0 to 0.5. and plot the numerically calculation with red solid line in Fig.\ref{fig:comparisonsetting}(d).

In contrast to previous work, the coherent control of channels scheme fails in our quantum dynamics tasks. This result is implicitly suggested by Eq.C3. The first terms in Eq.C3 represents that the system trivially being heated by the reservoir. While it is surprising that the population ratio between ground and excited state, which closely relates to energy or temperature, in the second term acts the same as initial system state, $\frac{ \mathcal{O} \rho \mathcal{O}^{\dagger}\left(2,2\right)}{\mathcal{O} \rho \mathcal{O}^{\dagger}\left(1,1\right)}=\frac{  \rho \left(2,2\right)}{ \rho \left(1,1\right)}$, which indicates no contribution to energy change.

However, historical hot debates compel cautious claim on the role of ICO in such quantum dynamics tasks. It was recently found that the advantages of ICO scheme over coherent control of channels depends on implementation of thermalizing channel \cite{wood2021operational}, e.g. usage of the control-swap operation introduced in ref.\cite{felce2020quantum} as thermalizing channel, where swapping of system qubit and reservoir qubit is accessible and controlled by ancillary qubit, both schemes give rise to quantum cooling effect \cite{nie2022quantum}. This suggests that ICO is not unique to realize quantum refrigeration with thermalizing channel, neither can we simply deny the the function of ICO as the it is still relevant even when for some Karus operators the coherent control of channels scheme failed, which is the case of our work. The coherent control of channels scheme is implementation-dependent while the ICO scheme is implementation-independent.  It is of interest to further investigate the role of ICO and figure out how a identical thermalizing channel with the different implementations may have different results. 
%We may guess that the non-markovianity is of relevance, since a single master equation would corresponds to various dynamics with different markovianity \cite{milz2019completely}. The generalized amplitude damping operators as thermalizing channel is deemed as markovian, while the swapping between system and reservoir qubit is non-markovian \cite{pollock2018operational}.    

{\it Channels in series with quantum control.---} In the second layout, we will analyze the system passing two trajectories in a superposition, where in each trajectory a fixed order of two channels is applied but different unitary operations between trajectories is allowed \cite{guerin2019communication}. The schematic is presented in Fig.\ref{fig:comparisonsetting}(c). For fair comparison, it is naturally assumed that the unitary operation should not solely alter the energy of system, or equivalently the population ratio between ground and excited state. This assumption could exclude the possibility of introducing additional energy to assist heat extraction. Consequently, the allowed unitary operation is deemed as a rotation along z-axis in Bloch sphere $U_{i}\left( \theta_{i}\right)=\left(\begin{array}{cc}
1 & 0\\
0 & e^{i\theta_{i}}
\end{array}\right)$ A complete description of channels in series with quantum control is provided as
\begin{eqnarray}
\mathcal{S}^{T}\left(\rho_{c}\otimes\rho\right)&=&\sum_{ij}M_{ij}\left(\rho_{c}\otimes\rho\right)M_{ij}^{\dagger}\\\nonumber
M_{ij}&=&\left|0\right\rangle \left\langle 0\right|_{c}K_{j}^{2}K_{i}^{1}+\left|1\right\rangle \left\langle 1\right|_{c}U_3\left(\theta_3 \right) K_{j}^{2}U_2\left(\theta_2 \right) K_{i}^{1}U_1\left(\theta_1 \right)
\end{eqnarray}
The resultant control-system joint state and postmeasured system state conditioned at projecting control qubit into $\left|\pm\right\rangle$ is respectively given by
\begin{eqnarray}
\mathcal{S}^{T}\left(\rho_{c}\otimes\rho\right) &= &\frac{1}{2} ( \left|0\right\rangle \left\langle 0\right|\otimes T+\left|1\right\rangle \left\langle 1\right|\otimes U_3 T U_3 ^{\dagger} \\\nonumber
& & +\left|0\right\rangle \left\langle 1\right| \mathrm{Tr}\left[ \rho U_1 ^{\dagger} \right] \mathrm{Tr}\left[ \rho U_2 ^{\dagger} \right] TU_3 ^{\dagger} +\left|1\right\rangle \left\langle 0\right| \mathrm{Tr}\left[  U_1 \rho \right] \mathrm{Tr}\left[ U_2  \rho \right] U_3 T )\\\nonumber
\rho_{\pm} & = & \frac{\left( T+ U_3 T U_3 ^{\dagger}\right) \pm \left( \mathrm{Tr}\left[ \rho U_1 ^{\dagger} \right] \mathrm{Tr}\left[ \rho U_2 ^{\dagger} \right] TU_3 ^{\dagger}+  \mathrm{Tr}\left[  U_1 \rho \right] \mathrm{Tr}\left[ U_2  \rho \right] U_3 T  \right)}{\mathrm{Tr} \left[ \left( T+ U_3 T U_3 ^{\dagger}\right) \pm \left( \mathrm{Tr}\left[ \rho U_1 ^{\dagger} \right] \mathrm{Tr}\left[ \rho U_2 ^{\dagger} \right] TU_3 ^{\dagger}+  \mathrm{Tr}\left[  U_1 \rho \right] \mathrm{Tr}\left[ U_2  \rho \right] U_3 T  \right) \right]}
\end{eqnarray}
Specifically, when the $U_3=\mathbb{I}$ is identity operator, arbitrary quantum operator for $U_1$ and $U_2$ end up in vain to non-classical heat extraction, the conditional output $\rho_{\pm}$ is always thermal state T. While the maximal non-classical heat extraction is accessible when we fix $U_1=U_2=\mathbb{I}$ and $U_3=\left(\begin{array}{cc}
1 & 0\\
0 & e^{i\pi}
\end{array}\right)$ introduce $\pi$ phase shift. The numerical calculation result is presented in Fig.\ref{fig:comparisonsetting}(d) with the green area representing the available range of weighted energy change $\triangle\widetilde{E}_{-}=p_{-}$ for arbitrary $U_i$ and the green solid line bounding the highest possible energy change by aforementioned setting.

In light of the foregoing analysis, we articulate precisely two main standpoints. 
%\sout{First, our numerical analysis illuminate the fact that although it is possible to transmit classical or quantum information through coherent superposition of quantum channel, this effort ends up in vain when it comes to our thermodynamical tasks. From a conceptual standpoint, different quantum features are relevant to different goals. For the quantum information transmission task, the quantum coherence carried by off-diagonal elements of system’s density matrix plays the central role. A superposed channel is to some extent able to preserve the superposed quantum state, and hence preserve quantum coherence. This ability is quantified by coherent information of channel or channel capacities} \cite{holevo1998capacity,schumacher1997sending,lloyd1997capacity,schumacher1996quantum}. \sout{However, in the context of thermal dynamical task, we focus on the thermal equilibrium state. The energy of system which corresponds to population of diagonal elements of density matrix plays an essential role. The perturbation in population is irrelevant to quantum coherence and channel capacity thus is substantially a new scene different from quantum communication tasks.}
First, for the thermalizing channel with specific kraus form in our work, imitation through coherent superposition of quantum channel ends up in vain for thermodynamical tasks. As the performance of coherent control of channel is implementation-dependent but the ICO scheme is implementation-independent, there should be more profound physical features behind which may help to deepen understanding of the role of ICO. It is also noticed that for information and thermodynamical task we both focus on the change of system state but in opposite perspectives. For the information transmission task, either in classical or quantum information transmission, we focus on to what extend the final state preserve the information of initial one. That is, the similarity between them. While in the context of thermal dynamical tasks, the heat extraction stems from the energy change which closely relates to deviation of final state with respect to initial one, thus is substantially a new scene.
%More apparently in comparison of transmission of quantum information and thermodynamical task, even different quantum features are relevant. The quantum coherence carried by off-diagonal elements of system’s density matrix plays the central role in quantum information transmission. A superposed channel is to some extent able to preserve the superposed quantum state, and hence preserve quantum coherence. This ability is quantified by coherent information \cite{lloyd1997capacity,schumacher1996quantum}. However, in thermodynamical task, we focus on the thermal equilibrium state. The energy of system which corresponds to population of diagonal elements of density matrix plays an essential role. The perturbation in population is irrelevant to quantum coherence and channel capacity thus is substantially a new scene.

The second standpoint is that it is possible to duplicate or even perform better in non-classical heat extraction by taking use of channels in series with quantum controlled operations. However, it should be noticed that an additional quantum controlled operation is required in this scheme . Such a quantum controlled operation requires the ability to manipulate the working substance microstate (basically it is control-unitary gate) rather than simply throw it into reservoir to reach thermal equilibrium. It is recognized that tracking and manipulating individual microscopic particle is such a strong postulate in thermodynamics that make the Maxwell's demon remain to be a gedanken experiment until recent decades \cite{maruyama2009colloquium,landauer1991information}. It is naturally expected that an extra equipped ability  will bring about a better performance than in ICO context.  It is hard to say manipulation on microscopic system is more resourceful than creating ICO as they are two independent protocols with different prerequisites. Our result suggests that the ICO can be an alternative for quantum refrigeration by using thermlizing channels soly and without manipulation on microscopic system state. Although in our optics experiment, the use of channels in series with quantum control use the same experiment resource as others, it is different from strictly theoretical viewpoint. 

\section{Further discussion on the indefinite causal order induced thermodynamics}

Most of the physical natures obey time-symmetric law where the system's evolutions do not intrinsically differentiate between forward and backward time directions \cite{rubino2020time}. But the second law of thermodynamics impose the irreversibility of thermodynamic time arrow on a closed system in which the entropy can never decrease. The second law bound the extractable work $W_{ext}<= \Delta F$ \cite{callen1985thermodynamics}, where $\Delta F$ is difference in the Helmholtz free energy between the initial and final thermodynamic equilibrium states. Regarding the case of system and reservoir sharing the same thermal state $T$ in our work, it is straightforward that simply thermal interaction between them will yield the output thermal state $T$. Hence the free energy can never increase under such case. The extractable work is doomed to be zero due to $W_{ext}=\Delta F=0$. This confirm the impossibility of extracting work out of a single reservoir according to Kelvin's statement \cite{pippard1964elements}.

\begin{figure}[ht]
\centering
\includegraphics[width=0.8\textwidth]{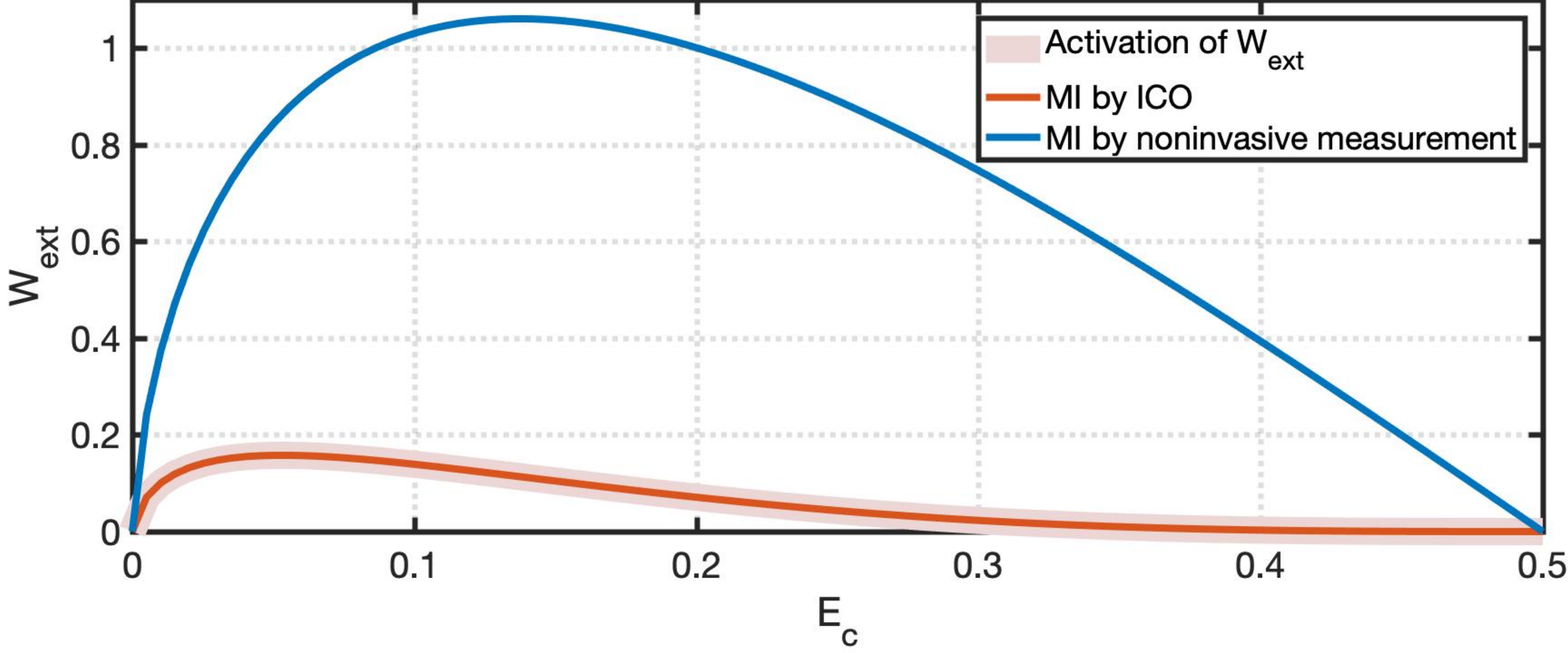}
\caption{\label{fig:maxwell} The work extraction ability of ICO and traditional Maxwell-demon scheme. In the case of ICO scenario, the theoretical calculated extractable work can be interpreted by activation of $W_{ext}$ (pink bold solid line) or equivalently information gain (red solid line) by measurement on ancillary control qubit. In the case of Maxwell demon scenario, the extractable work is represented by blue solid line. In both case, the initial system state's energy (horizontal coordinate) ranges from 0 to 0.5. Although the Maxwell demon have a better performance in extractable work, a more strict manipulation ability is imperative.}
\label{}
\end{figure}

However, when we extend the quantum mechanics into thermodynamics, the superposition law may enable the superposition of forward-in-time and backward-in-time process, which will lead to activation of extractable work. To be more detailed, the free energy of arbitrary quantum state $\rho$ with respect to a thermal reservoir with inverse temperature $\beta$ is defined as 
\begin{eqnarray}
F\left( \rho \right)=\mathrm{Tr}\left[\rho \mathcal{H}\right]-\beta \mathcal{S}\left(\rho\right)
\end{eqnarray}
where $\mathcal{S}\left(\rho\right)$ is the von Neumann entropy of system state $\rho$. The extractable work from the conditional output state is characterized as
\begin{eqnarray}
W_{ext}=\Delta F^{\prime}=p_{+}\left[F\left(\rho_{+}\right)-F\left(T\right)\right]+p_{-}\left[F\left(\rho_{-}\right)-F\left(T\right)\right]
\end{eqnarray}
where all the notions are the same with main text. The theoretical calculation with traversal of temperature (in terms of energy) is plotted by bold pink line in Fig.\ref{fig:maxwell}. A certain amount of extractable work is universally accessible except the exceptional point (E=0 and E=0.5, corresponding to temperature being zero and infinite). This analytical results suggests that two thermalizing channel in causally nonseparable order could lead to activation of extractable work. Our result indicates the possibility of certain thermodynamics tasks which only take use of thermalizing channels alone.

Such an activation of free energy difference can also be interpreted by distilling the information of system, which behaves in a manner of Maxwell-demon-like mechanism. It is interesting to reformulate in this way . Generalization of second law  of thermodynamics provides a reconciliation by including mutual information obtained by feedback control \cite{sagawa2010generalized,sagawa2008second},
\begin{eqnarray}
W_{ext} & = & \triangle F+\frac{1}{\beta}I\left(\rho_{1}\colon\Lambda\right)\\\nonumber
I\left(\rho_{1}\colon\Lambda\right) & = & \chi\left(\left\{ \rho_{2}^{\left(k\right)}\right\} \right)-\triangle S_{meas}
\end{eqnarray}
The $I\left(\rho_{1}\colon\Lambda\right)$ represents the information gain about the measured system $\rho_1$ by the measurement $\Lambda$, $\chi\left(\left\{ \rho_{2}^{\left(k\right)}\right\} \right)=S\left(\rho_{2}\right)-\sum_{k=\pm}p_{k}S\left(\rho_{2}^{\left(k\right)}\right)$ is the Holevo $\chi$ quantity, $\triangle S_{meas}=S\left(\rho_{2}\right)-S\left(\rho_{1}\right)$ is the von Neumann entropy difference of pre-measured $\rho_1=\mathrm{Tr}_{c}\left[\mathcal{S}^{T}\left(\rho_{c}\otimes\rho\right)\right]$and post-measured working system $\rho_2=\sum_{k=\pm}p_{k}\rho_{2}^{\left(k\right)}$ ($\rho_{2}^{\left(k\right)}$ is the conditional output conditioned at the measurement outcome $k$ of $\Lambda$) of which the expressions are provided in main text. The theoretical mutual information gain with the measurement being $\varLambda=\left\{ \left|+\right\rangle \left\langle +\right|_{c}\otimes I_{s},\left|-\right\rangle \left\langle -\right|_{c}\otimes I_{s}\right\}$ on control(c)-system(s) joint state is plotted in Fig.\ref{fig:maxwell} (red solid line), which overlaps with the aforementioned activation of extractable work (bold pink line), indicating the equivalence of these two interpretations.

It is interested to discuss the difference of our scheme with several previous Maxwell demons experiment. Maxwell demon steers the heat by adopting the feedback control, as well as acquisition of information about working substance by means of either invasive measurement or noninvasive measurement. Invasive measurement by directly interrogating on working system brings about the disturbance \cite{elouard2017extracting,elouard2018efficient,elouard2017role}, which is not the case of ours. The noninvasive measurement means that the measurement basis must coincide with the spectrum decomposition basis of working system state, indicating that in such basis the system state is diagonal \cite{koski2014experimental,koski2014experimentalreal}. A scheme with noninvasive measurement ( the measurement basis is energy basis $\varLambda=\left\{ \left|0\right\rangle \left\langle 0\right|_{s},\left|1\right\rangle \left\langle 1\right|_{s}\right\}$) on working system alone is also analyzed and the corresponding extractable work is plotted in Fig.\ref{fig:maxwell}. The underlying ideas of 
noninvasive schemes and ours are subtly different. Instead of directly probing the system, our scheme only resorts to probing the control qubit on fixed measurement basis, which is sort of indirect measurement as ref.\cite{quan2006maxwell,camati2016experimental}. Also the thermalizing channel as feedback control rather than manipulating individual microscopic particle gives a compromise of control ability on quantum system. There is still minor difference between our work and ref.\cite{quan2006maxwell,camati2016experimental} in corresponding operational process. Regarding their Maxwell-demon circuit, three typical steps sequentially take place per cycle, (i) Evolution. Working system evolves governed by specific Hamiltonian , (ii)  Coupling. Ability of coupling system-ancilla is necessary to transfer the information of system in to ancilla for indirect measurement (iii) Feedback. A control-unitary feedback operation on working system is  conditioned on the logic state of ancilla. The implementation of Maxwell demon with indirect measurement is based on coupling the system with ancilla. Given the thermalizing channel of our work, they can not get entangled by simply applying the channel. On the contrast, in the ICO scenario, the superposed time arrows naturally couples the system and ancillary control qubit in entangled trajectories , which exempts the coupling ability and simplify the process (i) Evolution. Working system evolves under indefinite causal order, (ii) Feedback. Thermalizing on working system is conditioned on the logic state of ancillary control qubit.

\section{Experimental setup for recycling photons}

\begin{figure}[ht]
\centering
\includegraphics[width=0.8\textwidth]{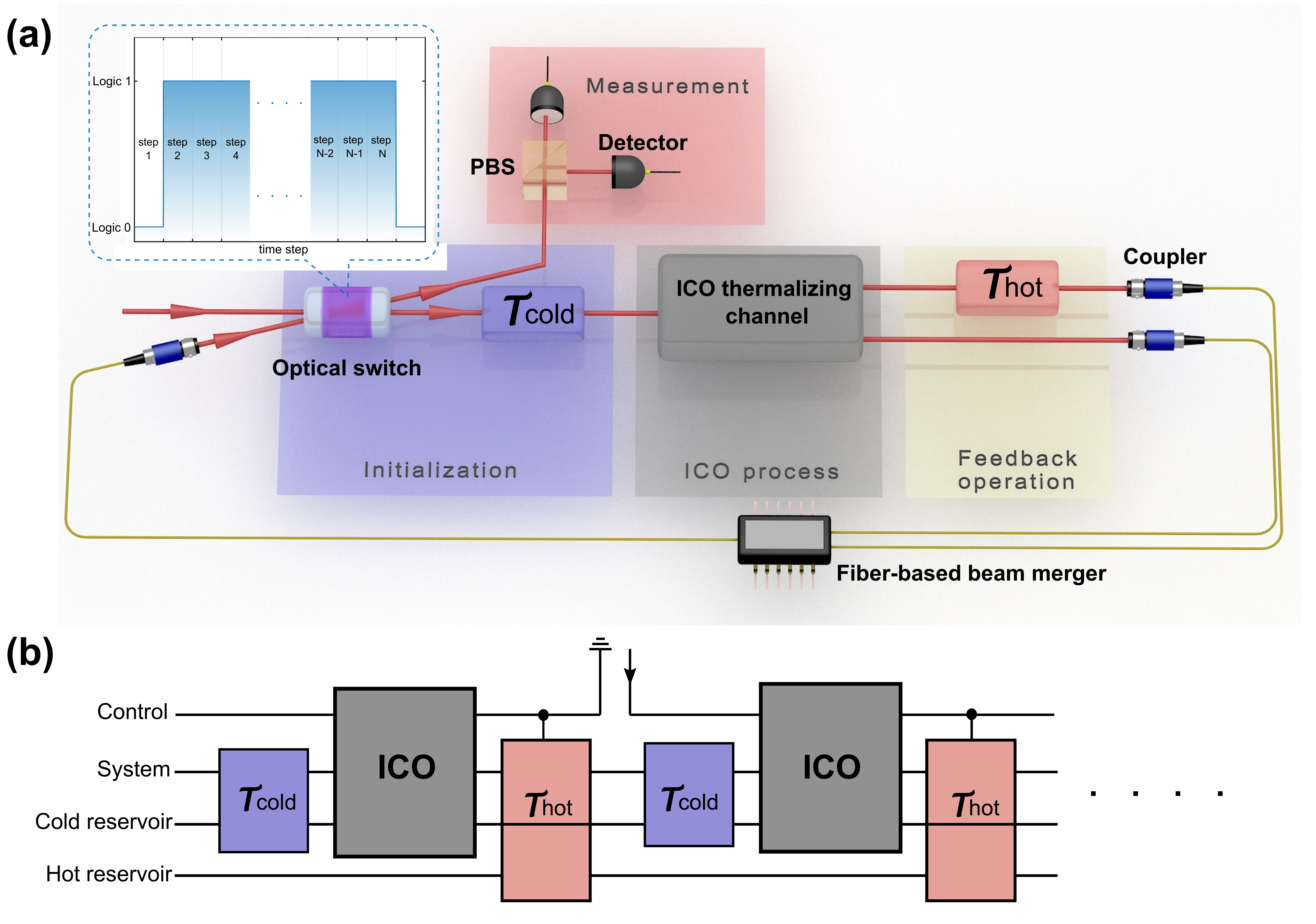}
\caption{(a) Extended setup for physically recycling the photon. Here the quantum channel is presented by the colored box. Blue (red) box: $\mathcal{T}_{cold}$($\mathcal{T}_{hot}$), thermal contact with cold (hot) reservoir. Grey box: thermal contact with reservoirs in indefinite causal order.The optical switch is driven by pulse signal with input of electric low-level signal (logic 0) corresponding to transmitting the photon and electric high-level signal (logic 1) corresponding to deflecting the photon.(b) Conceptual quantum circuit for the setup above. In every recycle time step, after the controlled-thermalizing channel with hot reservoir, the control qubit is discarded and a new one is introduced in ICO process. Consequently, the quantum refrigerator runs in consumption of purity of control qubit. }
\label{fig:recycle}
\end{figure}

To keep the quantum refrigerator running, the working substance should be recycled. In our photonic implementation, the photons are not physically recycled, but a new photon is prepared on control qubit. In this section, we heuristically introduce an extended setup for recycling photon. The extended setup implement the feedback operation in control-thermalizing channel without measuring the photon that will annihilates it. 

Sketch of the setup is drawn in Fig.\ref{fig:recycle}(a) and simply divided into four part: (1) Initialization. The working substance is initialized into thermal state of cold reservoir by thermally contacting with it. (2) ICO process. The initial working substance undergoes the thermal channel with indefinite causal order of which the setup based on quantum switch is already described in main text. (3) Feedback operation. A control-thermalizing channel is applied by thermal contact with external hot reservoir to release heat, conditioned on the control qubit in $\left|-\right\rangle$. After the feedback operation, two paths of photon are collected by couplers and combined by fiber-based beam merger. The optical switch can be switched on to deflect the recycled photon into initialization setup to start a new cycle, or switched off to stop recycling by sending the photon into measurement part. (4) Measurement. Measuring the population on ground and excited state.
The function optical switch can be realized by acoustic optical modulator (AOM). The modulation pulse is aforehand programmed. As is provided in the inset of Fig.\ref{fig:recycle}, one recycling period is denoted by one time step. In the first time step, the AOM is switched off to let the photon go into setup. In the following time step, the AOM is switched on to keep the photon recycling. In the final time step N, the AOM is switched off again to measure the final result. Because the AOM is driven by an aforehand programmed pulse, information about the system is not mandatory, thus the quantum refrigerator is fully functioned without measurement (annihilation of photon). In every cycle, the path qubit is discarded in fiber-based beam merger and a new path qubit is introduced in ICO process. Thus the quantum refrigerator functions without projection but rather with consumption of purity of control qubit. The conceptual quantum circuit is drawn in Fig.\ref{fig:recycle}(b).

\normalem
\bibliographystyle{apsrev}
\bibliography{arxivICOQR}

\begin{thebibliography}{84}
\expandafter\ifx\csname natexlab\endcsname\relax\def\natexlab#1{#1}\fi
\expandafter\ifx\csname bibnamefont\endcsname\relax
  \def\bibnamefont#1{#1}\fi
\expandafter\ifx\csname bibfnamefont\endcsname\relax
  \def\bibfnamefont#1{#1}\fi
\expandafter\ifx\csname citenamefont\endcsname\relax
  \def\citenamefont#1{#1}\fi
\expandafter\ifx\csname url\endcsname\relax
  \def\url#1{\texttt{#1}}\fi
\expandafter\ifx\csname urlprefix\endcsname\relax\def\urlprefix{URL }\fi
\providecommand{\bibinfo}[2]{#2}
\providecommand{\eprint}[2][]{\url{#2}}

\bibitem[{\citenamefont{Bell}(1964)}]{bell1964einstein}
\bibinfo{author}{\bibfnamefont{J.~S.} \bibnamefont{Bell}},
  \bibinfo{journal}{Physics Physique Fizika} \textbf{\bibinfo{volume}{1}},
  \bibinfo{pages}{195} (\bibinfo{year}{1964}).

\bibitem[{\citenamefont{Kochen and Specker}(1975)}]{kochen1975problem}
\bibinfo{author}{\bibfnamefont{S.}~\bibnamefont{Kochen}} \bibnamefont{and}
  \bibinfo{author}{\bibfnamefont{E.~P.} \bibnamefont{Specker}}, in
  \emph{\bibinfo{booktitle}{The logico-algebraic approach to quantum
  mechanics}} (\bibinfo{publisher}{Springer}, \bibinfo{year}{1975}), pp.
  \bibinfo{pages}{293--328}.

\bibitem[{\citenamefont{Oreshkov et~al.}(2012)\citenamefont{Oreshkov, Costa,
  and Brukner}}]{oreshkov2012quantum}
\bibinfo{author}{\bibfnamefont{O.}~\bibnamefont{Oreshkov}},
  \bibinfo{author}{\bibfnamefont{F.}~\bibnamefont{Costa}}, \bibnamefont{and}
  \bibinfo{author}{\bibfnamefont{{\v{C}}.}~\bibnamefont{Brukner}},
  \bibinfo{journal}{Nature communications} \textbf{\bibinfo{volume}{3}},
  \bibinfo{pages}{1} (\bibinfo{year}{2012}).

\bibitem[{\citenamefont{Brukner}(2014)}]{brukner2014quantum}
\bibinfo{author}{\bibfnamefont{{\v{C}}.}~\bibnamefont{Brukner}},
  \bibinfo{journal}{Nature Physics} \textbf{\bibinfo{volume}{10}},
  \bibinfo{pages}{259} (\bibinfo{year}{2014}).

\bibitem[{\citenamefont{Oriti}(2009)}]{oriti2009approaches}
\bibinfo{author}{\bibfnamefont{D.}~\bibnamefont{Oriti}},
  \emph{\bibinfo{title}{Approaches to quantum gravity: Toward a new
  understanding of space, time and matter}} (\bibinfo{publisher}{Cambridge
  University Press}, \bibinfo{year}{2009}).

\bibitem[{\citenamefont{Hossenfelder}(2017)}]{hossenfelder2017experimental}
\bibinfo{author}{\bibfnamefont{S.}~\bibnamefont{Hossenfelder}},
  \emph{\bibinfo{title}{Experimental search for quantum gravity}}
  (\bibinfo{publisher}{Springer}, \bibinfo{year}{2017}).

\bibitem[{\citenamefont{Marletto and
  Vedral}(2017)}]{marletto2017gravitationally}
\bibinfo{author}{\bibfnamefont{C.}~\bibnamefont{Marletto}} \bibnamefont{and}
  \bibinfo{author}{\bibfnamefont{V.}~\bibnamefont{Vedral}},
  \bibinfo{journal}{Physical review letters} \textbf{\bibinfo{volume}{119}},
  \bibinfo{pages}{240402} (\bibinfo{year}{2017}).

\bibitem[{\citenamefont{Bose et~al.}(2017)\citenamefont{Bose, Mazumdar, Morley,
  Ulbricht, Toro{\v{s}}, Paternostro, Geraci, Barker, Kim, and
  Milburn}}]{bose2017spin}
\bibinfo{author}{\bibfnamefont{S.}~\bibnamefont{Bose}},
  \bibinfo{author}{\bibfnamefont{A.}~\bibnamefont{Mazumdar}},
  \bibinfo{author}{\bibfnamefont{G.~W.} \bibnamefont{Morley}},
  \bibinfo{author}{\bibfnamefont{H.}~\bibnamefont{Ulbricht}},
  \bibinfo{author}{\bibfnamefont{M.}~\bibnamefont{Toro{\v{s}}}},
  \bibinfo{author}{\bibfnamefont{M.}~\bibnamefont{Paternostro}},
  \bibinfo{author}{\bibfnamefont{A.~A.} \bibnamefont{Geraci}},
  \bibinfo{author}{\bibfnamefont{P.~F.} \bibnamefont{Barker}},
  \bibinfo{author}{\bibfnamefont{M.~S.} \bibnamefont{Kim}}, \bibnamefont{and}
  \bibinfo{author}{\bibfnamefont{G.}~\bibnamefont{Milburn}},
  \bibinfo{journal}{Physical review letters} \textbf{\bibinfo{volume}{119}},
  \bibinfo{pages}{240401} (\bibinfo{year}{2017}).

\bibitem[{\citenamefont{Peres and Terno}(2004)}]{peres2004quantum}
\bibinfo{author}{\bibfnamefont{A.}~\bibnamefont{Peres}} \bibnamefont{and}
  \bibinfo{author}{\bibfnamefont{D.~R.} \bibnamefont{Terno}},
  \bibinfo{journal}{Reviews of Modern Physics} \textbf{\bibinfo{volume}{76}},
  \bibinfo{pages}{93} (\bibinfo{year}{2004}).

\bibitem[{\citenamefont{Christodoulou and
  Rovelli}(2019)}]{christodoulou2019possibility}
\bibinfo{author}{\bibfnamefont{M.}~\bibnamefont{Christodoulou}}
  \bibnamefont{and} \bibinfo{author}{\bibfnamefont{C.}~\bibnamefont{Rovelli}},
  \bibinfo{journal}{Physics Letters B} \textbf{\bibinfo{volume}{792}},
  \bibinfo{pages}{64} (\bibinfo{year}{2019}).

\bibitem[{\citenamefont{DeWitt}(1967)}]{dewitt1967quantum}
\bibinfo{author}{\bibfnamefont{B.~S.} \bibnamefont{DeWitt}},
  \bibinfo{journal}{Physical Review} \textbf{\bibinfo{volume}{160}},
  \bibinfo{pages}{1113} (\bibinfo{year}{1967}).

\bibitem[{\citenamefont{Rovelli}(1990)}]{rovelli1990quantum}
\bibinfo{author}{\bibfnamefont{C.}~\bibnamefont{Rovelli}},
  \bibinfo{journal}{Physical Review D} \textbf{\bibinfo{volume}{42}},
  \bibinfo{pages}{2638} (\bibinfo{year}{1990}).

\bibitem[{\citenamefont{Gambini et~al.}(2004)\citenamefont{Gambini, Porto, and
  Pullin}}]{gambini2004relational}
\bibinfo{author}{\bibfnamefont{R.}~\bibnamefont{Gambini}},
  \bibinfo{author}{\bibfnamefont{R.~A.} \bibnamefont{Porto}}, \bibnamefont{and}
  \bibinfo{author}{\bibfnamefont{J.}~\bibnamefont{Pullin}},
  \bibinfo{journal}{New Journal of Physics} \textbf{\bibinfo{volume}{6}},
  \bibinfo{pages}{45} (\bibinfo{year}{2004}).

\bibitem[{\citenamefont{Hardy}(2009)}]{hardy2009quantum}
\bibinfo{author}{\bibfnamefont{L.}~\bibnamefont{Hardy}}, in
  \emph{\bibinfo{booktitle}{Quantum reality, relativistic causality, and
  closing the epistemic circle}} (\bibinfo{publisher}{Springer},
  \bibinfo{year}{2009}), pp. \bibinfo{pages}{379--401}.

\bibitem[{\citenamefont{Taddei et~al.}(2019)\citenamefont{Taddei, Nery, and
  Aolita}}]{taddei2019quantum}
\bibinfo{author}{\bibfnamefont{M.~M.} \bibnamefont{Taddei}},
  \bibinfo{author}{\bibfnamefont{R.~V.} \bibnamefont{Nery}}, \bibnamefont{and}
  \bibinfo{author}{\bibfnamefont{L.}~\bibnamefont{Aolita}},
  \bibinfo{journal}{Physical Review Research} \textbf{\bibinfo{volume}{1}},
  \bibinfo{pages}{033174} (\bibinfo{year}{2019}).

\bibitem[{\citenamefont{Jia and Costa}(2019)}]{jia2019causal}
\bibinfo{author}{\bibfnamefont{D.}~\bibnamefont{Jia}} \bibnamefont{and}
  \bibinfo{author}{\bibfnamefont{F.}~\bibnamefont{Costa}},
  \bibinfo{journal}{Physical Review A} \textbf{\bibinfo{volume}{100}},
  \bibinfo{pages}{052319} (\bibinfo{year}{2019}).

\bibitem[{\citenamefont{Chiribella}(2012)}]{chiribella2012perfect}
\bibinfo{author}{\bibfnamefont{G.}~\bibnamefont{Chiribella}},
  \bibinfo{journal}{Physical Review A} \textbf{\bibinfo{volume}{86}},
  \bibinfo{pages}{040301} (\bibinfo{year}{2012}).

\bibitem[{\citenamefont{Gu{\'e}rin et~al.}(2016)\citenamefont{Gu{\'e}rin, Feix,
  Ara{\'u}jo, and Brukner}}]{guerin2016exponential}
\bibinfo{author}{\bibfnamefont{P.~A.} \bibnamefont{Gu{\'e}rin}},
  \bibinfo{author}{\bibfnamefont{A.}~\bibnamefont{Feix}},
  \bibinfo{author}{\bibfnamefont{M.}~\bibnamefont{Ara{\'u}jo}},
  \bibnamefont{and}
  \bibinfo{author}{\bibfnamefont{{\v{C}}.}~\bibnamefont{Brukner}},
  \bibinfo{journal}{Physical review letters} \textbf{\bibinfo{volume}{117}},
  \bibinfo{pages}{100502} (\bibinfo{year}{2016}).

\bibitem[{\citenamefont{Feix et~al.}(2015)\citenamefont{Feix, Ara{\'u}jo, and
  Brukner}}]{feix2015quantum}
\bibinfo{author}{\bibfnamefont{A.}~\bibnamefont{Feix}},
  \bibinfo{author}{\bibfnamefont{M.}~\bibnamefont{Ara{\'u}jo}},
  \bibnamefont{and}
  \bibinfo{author}{\bibfnamefont{{\v{C}}.}~\bibnamefont{Brukner}},
  \bibinfo{journal}{Physical Review A} \textbf{\bibinfo{volume}{92}},
  \bibinfo{pages}{052326} (\bibinfo{year}{2015}).

\bibitem[{\citenamefont{Ara{\'u}jo
  et~al.}(2014{\natexlab{a}})\citenamefont{Ara{\'u}jo, Costa, and
  Brukner}}]{araujo2014computational}
\bibinfo{author}{\bibfnamefont{M.}~\bibnamefont{Ara{\'u}jo}},
  \bibinfo{author}{\bibfnamefont{F.}~\bibnamefont{Costa}}, \bibnamefont{and}
  \bibinfo{author}{\bibfnamefont{{\v{C}}.}~\bibnamefont{Brukner}},
  \bibinfo{journal}{Physical review letters} \textbf{\bibinfo{volume}{113}},
  \bibinfo{pages}{250402} (\bibinfo{year}{2014}{\natexlab{a}}).

\bibitem[{\citenamefont{Zhao et~al.}(2020)\citenamefont{Zhao, Yang, and
  Chiribella}}]{zhao2020quantum}
\bibinfo{author}{\bibfnamefont{X.}~\bibnamefont{Zhao}},
  \bibinfo{author}{\bibfnamefont{Y.}~\bibnamefont{Yang}}, \bibnamefont{and}
  \bibinfo{author}{\bibfnamefont{G.}~\bibnamefont{Chiribella}},
  \bibinfo{journal}{Physical Review Letters} \textbf{\bibinfo{volume}{124}},
  \bibinfo{pages}{190503} (\bibinfo{year}{2020}).

\bibitem[{\citenamefont{Mukhopadhyay et~al.}(2018)\citenamefont{Mukhopadhyay,
  Gupta, and Pati}}]{mukhopadhyay2018superposition}
\bibinfo{author}{\bibfnamefont{C.}~\bibnamefont{Mukhopadhyay}},
  \bibinfo{author}{\bibfnamefont{M.~K.} \bibnamefont{Gupta}}, \bibnamefont{and}
  \bibinfo{author}{\bibfnamefont{A.~K.} \bibnamefont{Pati}},
  \bibinfo{journal}{arXiv preprint arXiv:1812.07508}  (\bibinfo{year}{2018}).

\bibitem[{\citenamefont{Ebler et~al.}(2018)\citenamefont{Ebler, Salek, and
  Chiribella}}]{ebler2018enhanced}
\bibinfo{author}{\bibfnamefont{D.}~\bibnamefont{Ebler}},
  \bibinfo{author}{\bibfnamefont{S.}~\bibnamefont{Salek}}, \bibnamefont{and}
  \bibinfo{author}{\bibfnamefont{G.}~\bibnamefont{Chiribella}},
  \bibinfo{journal}{Physical review letters} \textbf{\bibinfo{volume}{120}},
  \bibinfo{pages}{120502} (\bibinfo{year}{2018}).

\bibitem[{\citenamefont{Salek et~al.}(2018)\citenamefont{Salek, Ebler, and
  Chiribella}}]{salek2018quantum}
\bibinfo{author}{\bibfnamefont{S.}~\bibnamefont{Salek}},
  \bibinfo{author}{\bibfnamefont{D.}~\bibnamefont{Ebler}}, \bibnamefont{and}
  \bibinfo{author}{\bibfnamefont{G.}~\bibnamefont{Chiribella}},
  \bibinfo{journal}{arXiv preprint arXiv:1809.06655}  (\bibinfo{year}{2018}).

\bibitem[{\citenamefont{Chiribella et~al.}(2018)\citenamefont{Chiribella,
  Banik, Bhattacharya, Guha, Alimuddin, Roy, Saha, Agrawal, and
  Kar}}]{chiribella2018indefinite}
\bibinfo{author}{\bibfnamefont{G.}~\bibnamefont{Chiribella}},
  \bibinfo{author}{\bibfnamefont{M.}~\bibnamefont{Banik}},
  \bibinfo{author}{\bibfnamefont{S.~S.} \bibnamefont{Bhattacharya}},
  \bibinfo{author}{\bibfnamefont{T.}~\bibnamefont{Guha}},
  \bibinfo{author}{\bibfnamefont{M.}~\bibnamefont{Alimuddin}},
  \bibinfo{author}{\bibfnamefont{A.}~\bibnamefont{Roy}},
  \bibinfo{author}{\bibfnamefont{S.}~\bibnamefont{Saha}},
  \bibinfo{author}{\bibfnamefont{S.}~\bibnamefont{Agrawal}}, \bibnamefont{and}
  \bibinfo{author}{\bibfnamefont{G.}~\bibnamefont{Kar}},
  \bibinfo{journal}{arXiv preprint arXiv:1810.10457}  (\bibinfo{year}{2018}).

\bibitem[{\citenamefont{Procopio et~al.}(2015)\citenamefont{Procopio, Moqanaki,
  Ara{\'u}jo, Costa, Calafell, Dowd, Hamel, Rozema, Brukner, and
  Walther}}]{procopio2015experimental}
\bibinfo{author}{\bibfnamefont{L.~M.} \bibnamefont{Procopio}},
  \bibinfo{author}{\bibfnamefont{A.}~\bibnamefont{Moqanaki}},
  \bibinfo{author}{\bibfnamefont{M.}~\bibnamefont{Ara{\'u}jo}},
  \bibinfo{author}{\bibfnamefont{F.}~\bibnamefont{Costa}},
  \bibinfo{author}{\bibfnamefont{I.~A.} \bibnamefont{Calafell}},
  \bibinfo{author}{\bibfnamefont{E.~G.} \bibnamefont{Dowd}},
  \bibinfo{author}{\bibfnamefont{D.~R.} \bibnamefont{Hamel}},
  \bibinfo{author}{\bibfnamefont{L.~A.} \bibnamefont{Rozema}},
  \bibinfo{author}{\bibfnamefont{{\v{C}}.}~\bibnamefont{Brukner}},
  \bibnamefont{and} \bibinfo{author}{\bibfnamefont{P.}~\bibnamefont{Walther}},
  \bibinfo{journal}{Nature communications} \textbf{\bibinfo{volume}{6}},
  \bibinfo{pages}{1} (\bibinfo{year}{2015}).

\bibitem[{\citenamefont{Rubino et~al.}(2017)\citenamefont{Rubino, Rozema, Feix,
  Ara{\'u}jo, Zeuner, Procopio, Brukner, and Walther}}]{rubino2017experimental}
\bibinfo{author}{\bibfnamefont{G.}~\bibnamefont{Rubino}},
  \bibinfo{author}{\bibfnamefont{L.~A.} \bibnamefont{Rozema}},
  \bibinfo{author}{\bibfnamefont{A.}~\bibnamefont{Feix}},
  \bibinfo{author}{\bibfnamefont{M.}~\bibnamefont{Ara{\'u}jo}},
  \bibinfo{author}{\bibfnamefont{J.~M.} \bibnamefont{Zeuner}},
  \bibinfo{author}{\bibfnamefont{L.~M.} \bibnamefont{Procopio}},
  \bibinfo{author}{\bibfnamefont{{\v{C}}.}~\bibnamefont{Brukner}},
  \bibnamefont{and} \bibinfo{author}{\bibfnamefont{P.}~\bibnamefont{Walther}},
  \bibinfo{journal}{Science advances} \textbf{\bibinfo{volume}{3}},
  \bibinfo{pages}{e1602589} (\bibinfo{year}{2017}).

\bibitem[{\citenamefont{Goswami et~al.}(2018)\citenamefont{Goswami, Giarmatzi,
  Kewming, Costa, Branciard, Romero, and White}}]{goswami2018indefinite}
\bibinfo{author}{\bibfnamefont{K.}~\bibnamefont{Goswami}},
  \bibinfo{author}{\bibfnamefont{C.}~\bibnamefont{Giarmatzi}},
  \bibinfo{author}{\bibfnamefont{M.}~\bibnamefont{Kewming}},
  \bibinfo{author}{\bibfnamefont{F.}~\bibnamefont{Costa}},
  \bibinfo{author}{\bibfnamefont{C.}~\bibnamefont{Branciard}},
  \bibinfo{author}{\bibfnamefont{J.}~\bibnamefont{Romero}}, \bibnamefont{and}
  \bibinfo{author}{\bibfnamefont{A.~G.} \bibnamefont{White}},
  \bibinfo{journal}{Physical Review Letters} \textbf{\bibinfo{volume}{121}},
  \bibinfo{pages}{090503} (\bibinfo{year}{2018}).

\bibitem[{\citenamefont{Guo et~al.}(2020)\citenamefont{Guo, Hu, Hou, Cao, Cui,
  Liu, Huang, Li, Guo, and Chiribella}}]{guo2020experimental}
\bibinfo{author}{\bibfnamefont{Y.}~\bibnamefont{Guo}},
  \bibinfo{author}{\bibfnamefont{X.-M.} \bibnamefont{Hu}},
  \bibinfo{author}{\bibfnamefont{Z.-B.} \bibnamefont{Hou}},
  \bibinfo{author}{\bibfnamefont{H.}~\bibnamefont{Cao}},
  \bibinfo{author}{\bibfnamefont{J.-M.} \bibnamefont{Cui}},
  \bibinfo{author}{\bibfnamefont{B.-H.} \bibnamefont{Liu}},
  \bibinfo{author}{\bibfnamefont{Y.-F.} \bibnamefont{Huang}},
  \bibinfo{author}{\bibfnamefont{C.-F.} \bibnamefont{Li}},
  \bibinfo{author}{\bibfnamefont{G.-C.} \bibnamefont{Guo}}, \bibnamefont{and}
  \bibinfo{author}{\bibfnamefont{G.}~\bibnamefont{Chiribella}},
  \bibinfo{journal}{Physical Review Letters} \textbf{\bibinfo{volume}{124}},
  \bibinfo{pages}{030502} (\bibinfo{year}{2020}).

\bibitem[{\citenamefont{Goswami et~al.}(2020)\citenamefont{Goswami, Cao,
  Paz-Silva, Romero, and White}}]{goswami2020increasing}
\bibinfo{author}{\bibfnamefont{K.}~\bibnamefont{Goswami}},
  \bibinfo{author}{\bibfnamefont{Y.}~\bibnamefont{Cao}},
  \bibinfo{author}{\bibfnamefont{G.~A.} \bibnamefont{Paz-Silva}},
  \bibinfo{author}{\bibfnamefont{J.}~\bibnamefont{Romero}}, \bibnamefont{and}
  \bibinfo{author}{\bibfnamefont{A.~G.} \bibnamefont{White}},
  \bibinfo{journal}{Physical Review Research} \textbf{\bibinfo{volume}{2}},
  \bibinfo{pages}{033292} (\bibinfo{year}{2020}).

\bibitem[{\citenamefont{Wei et~al.}(2019)\citenamefont{Wei, Tischler, Zhao, Li,
  Arrazola, Liu, Zhang, Li, You, Wang et~al.}}]{wei2019experimental}
\bibinfo{author}{\bibfnamefont{K.}~\bibnamefont{Wei}},
  \bibinfo{author}{\bibfnamefont{N.}~\bibnamefont{Tischler}},
  \bibinfo{author}{\bibfnamefont{S.-R.} \bibnamefont{Zhao}},
  \bibinfo{author}{\bibfnamefont{Y.-H.} \bibnamefont{Li}},
  \bibinfo{author}{\bibfnamefont{J.~M.} \bibnamefont{Arrazola}},
  \bibinfo{author}{\bibfnamefont{Y.}~\bibnamefont{Liu}},
  \bibinfo{author}{\bibfnamefont{W.}~\bibnamefont{Zhang}},
  \bibinfo{author}{\bibfnamefont{H.}~\bibnamefont{Li}},
  \bibinfo{author}{\bibfnamefont{L.}~\bibnamefont{You}},
  \bibinfo{author}{\bibfnamefont{Z.}~\bibnamefont{Wang}}, \bibnamefont{et~al.},
  \bibinfo{journal}{Phys. Rev. Lett.} \textbf{\bibinfo{volume}{122}},
  \bibinfo{pages}{120504} (\bibinfo{year}{2019}).

\bibitem[{\citenamefont{Felce and Vedral}(2020)}]{felce2020quantum}
\bibinfo{author}{\bibfnamefont{D.}~\bibnamefont{Felce}} \bibnamefont{and}
  \bibinfo{author}{\bibfnamefont{V.}~\bibnamefont{Vedral}},
  \bibinfo{journal}{Physical Review Letters} \textbf{\bibinfo{volume}{125}},
  \bibinfo{pages}{070603} (\bibinfo{year}{2020}).

\bibitem[{\citenamefont{Campisi et~al.}(2015)\citenamefont{Campisi, Pekola, and
  Fazio}}]{campisi2015nonequilibrium}
\bibinfo{author}{\bibfnamefont{M.}~\bibnamefont{Campisi}},
  \bibinfo{author}{\bibfnamefont{J.}~\bibnamefont{Pekola}}, \bibnamefont{and}
  \bibinfo{author}{\bibfnamefont{R.}~\bibnamefont{Fazio}},
  \bibinfo{journal}{New Journal of Physics} \textbf{\bibinfo{volume}{17}},
  \bibinfo{pages}{035012} (\bibinfo{year}{2015}).

\bibitem[{\citenamefont{Campisi and Fazio}(2016)}]{campisi2016dissipation}
\bibinfo{author}{\bibfnamefont{M.}~\bibnamefont{Campisi}} \bibnamefont{and}
  \bibinfo{author}{\bibfnamefont{R.}~\bibnamefont{Fazio}},
  \bibinfo{journal}{Journal of Physics A: Mathematical and Theoretical}
  \textbf{\bibinfo{volume}{49}}, \bibinfo{pages}{345002}
  (\bibinfo{year}{2016}).

\bibitem[{\citenamefont{Maruyama et~al.}(2009)\citenamefont{Maruyama, Nori, and
  Vedral}}]{maruyama2009colloquium}
\bibinfo{author}{\bibfnamefont{K.}~\bibnamefont{Maruyama}},
  \bibinfo{author}{\bibfnamefont{F.}~\bibnamefont{Nori}}, \bibnamefont{and}
  \bibinfo{author}{\bibfnamefont{V.}~\bibnamefont{Vedral}},
  \bibinfo{journal}{Reviews of Modern Physics} \textbf{\bibinfo{volume}{81}},
  \bibinfo{pages}{1} (\bibinfo{year}{2009}).

\bibitem[{\citenamefont{Elouard
  et~al.}(2017{\natexlab{a}})\citenamefont{Elouard, Herrera-Mart{\'\i}, Huard,
  and Auffeves}}]{elouard2017extracting}
\bibinfo{author}{\bibfnamefont{C.}~\bibnamefont{Elouard}},
  \bibinfo{author}{\bibfnamefont{D.}~\bibnamefont{Herrera-Mart{\'\i}}},
  \bibinfo{author}{\bibfnamefont{B.}~\bibnamefont{Huard}}, \bibnamefont{and}
  \bibinfo{author}{\bibfnamefont{A.}~\bibnamefont{Auffeves}},
  \bibinfo{journal}{Physical Review Letters} \textbf{\bibinfo{volume}{118}},
  \bibinfo{pages}{260603} (\bibinfo{year}{2017}{\natexlab{a}}).

\bibitem[{\citenamefont{Buffoni et~al.}(2019)\citenamefont{Buffoni, Solfanelli,
  Verrucchi, Cuccoli, and Campisi}}]{buffoni2019quantum}
\bibinfo{author}{\bibfnamefont{L.}~\bibnamefont{Buffoni}},
  \bibinfo{author}{\bibfnamefont{A.}~\bibnamefont{Solfanelli}},
  \bibinfo{author}{\bibfnamefont{P.}~\bibnamefont{Verrucchi}},
  \bibinfo{author}{\bibfnamefont{A.}~\bibnamefont{Cuccoli}}, \bibnamefont{and}
  \bibinfo{author}{\bibfnamefont{M.}~\bibnamefont{Campisi}},
  \bibinfo{journal}{Physical review letters} \textbf{\bibinfo{volume}{122}},
  \bibinfo{pages}{070603} (\bibinfo{year}{2019}).

\bibitem[{\citenamefont{Guha et~al.}(2020)\citenamefont{Guha, Alimuddin, and
  Parashar}}]{guha2020thermodynamic}
\bibinfo{author}{\bibfnamefont{T.}~\bibnamefont{Guha}},
  \bibinfo{author}{\bibfnamefont{M.}~\bibnamefont{Alimuddin}},
  \bibnamefont{and} \bibinfo{author}{\bibfnamefont{P.}~\bibnamefont{Parashar}},
  \bibinfo{journal}{arXiv preprint arXiv:2003.01464}  (\bibinfo{year}{2020}).

\bibitem[{\citenamefont{Markes and Hardy}(2011)}]{markes2011entropy}
\bibinfo{author}{\bibfnamefont{S.}~\bibnamefont{Markes}} \bibnamefont{and}
  \bibinfo{author}{\bibfnamefont{L.}~\bibnamefont{Hardy}}, in
  \emph{\bibinfo{booktitle}{Journal of Physics-Conference Series}}
  (\bibinfo{year}{2011}), vol. \bibinfo{volume}{306}, p.
  \bibinfo{pages}{012043}.

\bibitem[{\citenamefont{Simonov et~al.}(2020)\citenamefont{Simonov, Francica,
  Guarnieri, and Paternostro}}]{simonov2020ergotropy}
\bibinfo{author}{\bibfnamefont{K.}~\bibnamefont{Simonov}},
  \bibinfo{author}{\bibfnamefont{G.}~\bibnamefont{Francica}},
  \bibinfo{author}{\bibfnamefont{G.}~\bibnamefont{Guarnieri}},
  \bibnamefont{and}
  \bibinfo{author}{\bibfnamefont{M.}~\bibnamefont{Paternostro}},
  \bibinfo{journal}{arXiv preprint arXiv:2009.11265}  (\bibinfo{year}{2020}).

\bibitem[{\citenamefont{Rubino et~al.}(2020)\citenamefont{Rubino, Manzano, and
  Brukner}}]{rubino2020time}
\bibinfo{author}{\bibfnamefont{G.}~\bibnamefont{Rubino}},
  \bibinfo{author}{\bibfnamefont{G.}~\bibnamefont{Manzano}}, \bibnamefont{and}
  \bibinfo{author}{\bibfnamefont{{\v{C}}.}~\bibnamefont{Brukner}},
  \bibinfo{journal}{arXiv preprint arXiv:2008.02818}  (\bibinfo{year}{2020}).

\bibitem[{\citenamefont{Chiribella et~al.}(2013)\citenamefont{Chiribella,
  D’Ariano, Perinotti, and Valiron}}]{chiribella2013quantum}
\bibinfo{author}{\bibfnamefont{G.}~\bibnamefont{Chiribella}},
  \bibinfo{author}{\bibfnamefont{G.~M.} \bibnamefont{D’Ariano}},
  \bibinfo{author}{\bibfnamefont{P.}~\bibnamefont{Perinotti}},
  \bibnamefont{and} \bibinfo{author}{\bibfnamefont{B.}~\bibnamefont{Valiron}},
  \bibinfo{journal}{Physical Review A} \textbf{\bibinfo{volume}{88}},
  \bibinfo{pages}{022318} (\bibinfo{year}{2013}).

\bibitem[{\citenamefont{Xu et~al.}(2014)\citenamefont{Xu, Yung, Xu, Boixo,
  Zhou, Li, Aspuru-Guzik, and Guo}}]{xu2014demon}
\bibinfo{author}{\bibfnamefont{J.-S.} \bibnamefont{Xu}},
  \bibinfo{author}{\bibfnamefont{M.-H.} \bibnamefont{Yung}},
  \bibinfo{author}{\bibfnamefont{X.-Y.} \bibnamefont{Xu}},
  \bibinfo{author}{\bibfnamefont{S.}~\bibnamefont{Boixo}},
  \bibinfo{author}{\bibfnamefont{Z.-W.} \bibnamefont{Zhou}},
  \bibinfo{author}{\bibfnamefont{C.-F.} \bibnamefont{Li}},
  \bibinfo{author}{\bibfnamefont{A.}~\bibnamefont{Aspuru-Guzik}},
  \bibnamefont{and} \bibinfo{author}{\bibfnamefont{G.-C.} \bibnamefont{Guo}},
  \bibinfo{journal}{Nature Photonics} \textbf{\bibinfo{volume}{8}},
  \bibinfo{pages}{113} (\bibinfo{year}{2014}).

\bibitem[{\citenamefont{Mancino et~al.}(2017)\citenamefont{Mancino, Sbroscia,
  Gianani, Roccia, and Barbieri}}]{mancino2017quantum}
\bibinfo{author}{\bibfnamefont{L.}~\bibnamefont{Mancino}},
  \bibinfo{author}{\bibfnamefont{M.}~\bibnamefont{Sbroscia}},
  \bibinfo{author}{\bibfnamefont{I.}~\bibnamefont{Gianani}},
  \bibinfo{author}{\bibfnamefont{E.}~\bibnamefont{Roccia}}, \bibnamefont{and}
  \bibinfo{author}{\bibfnamefont{M.}~\bibnamefont{Barbieri}},
  \bibinfo{journal}{Physical Review Letters} \textbf{\bibinfo{volume}{118}},
  \bibinfo{pages}{130502} (\bibinfo{year}{2017}).

\bibitem[{\citenamefont{Fisher et~al.}(2012)\citenamefont{Fisher, Prevedel,
  Kaltenbaek, and Resch}}]{fisher2012optimal}
\bibinfo{author}{\bibfnamefont{K.~A.} \bibnamefont{Fisher}},
  \bibinfo{author}{\bibfnamefont{R.}~\bibnamefont{Prevedel}},
  \bibinfo{author}{\bibfnamefont{R.}~\bibnamefont{Kaltenbaek}},
  \bibnamefont{and} \bibinfo{author}{\bibfnamefont{K.~J.} \bibnamefont{Resch}},
  \bibinfo{journal}{New Journal of Physics} \textbf{\bibinfo{volume}{14}},
  \bibinfo{pages}{033016} (\bibinfo{year}{2012}).

\bibitem[{\citenamefont{Lu et~al.}(2017)\citenamefont{Lu, Liu, Wang, Chen, Li,
  Yao, Li, Liu, Peng, Sanders et~al.}}]{lu2017experimental}
\bibinfo{author}{\bibfnamefont{H.}~\bibnamefont{Lu}},
  \bibinfo{author}{\bibfnamefont{C.}~\bibnamefont{Liu}},
  \bibinfo{author}{\bibfnamefont{D.-S.} \bibnamefont{Wang}},
  \bibinfo{author}{\bibfnamefont{L.-K.} \bibnamefont{Chen}},
  \bibinfo{author}{\bibfnamefont{Z.-D.} \bibnamefont{Li}},
  \bibinfo{author}{\bibfnamefont{X.-C.} \bibnamefont{Yao}},
  \bibinfo{author}{\bibfnamefont{L.}~\bibnamefont{Li}},
  \bibinfo{author}{\bibfnamefont{N.-L.} \bibnamefont{Liu}},
  \bibinfo{author}{\bibfnamefont{C.-Z.} \bibnamefont{Peng}},
  \bibinfo{author}{\bibfnamefont{B.~C.} \bibnamefont{Sanders}},
  \bibnamefont{et~al.}, \bibinfo{journal}{Physical Review A}
  \textbf{\bibinfo{volume}{95}}, \bibinfo{pages}{042310}
  (\bibinfo{year}{2017}).

\bibitem[{\citenamefont{Pippard}(1964)}]{pippard1964elements}
\bibinfo{author}{\bibfnamefont{A.~B.} \bibnamefont{Pippard}},
  \emph{\bibinfo{title}{Elements of classical thermodynamics: for advanced
  students of physics}} (\bibinfo{publisher}{Cambridge University Press},
  \bibinfo{year}{1964}).

\bibitem[{\citenamefont{Abdelkhalek et~al.}(2016)\citenamefont{Abdelkhalek,
  Nakata, and Reeb}}]{abdelkhalek2016fundamental}
\bibinfo{author}{\bibfnamefont{K.}~\bibnamefont{Abdelkhalek}},
  \bibinfo{author}{\bibfnamefont{Y.}~\bibnamefont{Nakata}}, \bibnamefont{and}
  \bibinfo{author}{\bibfnamefont{D.}~\bibnamefont{Reeb}},
  \bibinfo{journal}{arXiv preprint arXiv:1609.06981}  (\bibinfo{year}{2016}).

\bibitem[{\citenamefont{Landauer}(1961)}]{landauer1961irreversibility}
\bibinfo{author}{\bibfnamefont{R.}~\bibnamefont{Landauer}},
  \bibinfo{journal}{IBM journal of research and development}
  \textbf{\bibinfo{volume}{5}}, \bibinfo{pages}{183} (\bibinfo{year}{1961}).

\bibitem[{\citenamefont{Perarnau-Llobet
  et~al.}(2015)\citenamefont{Perarnau-Llobet, Hovhannisyan, Huber, Skrzypczyk,
  Brunner, and Ac{\i}n}}]{perarnau2015extracting}
\bibinfo{author}{\bibfnamefont{M.}~\bibnamefont{Perarnau-Llobet}},
  \bibinfo{author}{\bibfnamefont{K.~V.} \bibnamefont{Hovhannisyan}},
  \bibinfo{author}{\bibfnamefont{M.}~\bibnamefont{Huber}},
  \bibinfo{author}{\bibfnamefont{P.}~\bibnamefont{Skrzypczyk}},
  \bibinfo{author}{\bibfnamefont{N.}~\bibnamefont{Brunner}}, \bibnamefont{and}
  \bibinfo{author}{\bibfnamefont{A.}~\bibnamefont{Ac{\i}n}},
  \bibinfo{journal}{Phys. Rev}  (\bibinfo{year}{2015}).

\bibitem[{\citenamefont{Francica et~al.}(2017)\citenamefont{Francica, Goold,
  Plastina, and Paternostro}}]{francica2017daemonic}
\bibinfo{author}{\bibfnamefont{G.}~\bibnamefont{Francica}},
  \bibinfo{author}{\bibfnamefont{J.}~\bibnamefont{Goold}},
  \bibinfo{author}{\bibfnamefont{F.}~\bibnamefont{Plastina}}, \bibnamefont{and}
  \bibinfo{author}{\bibfnamefont{M.}~\bibnamefont{Paternostro}},
  \bibinfo{journal}{npj Quantum Information} \textbf{\bibinfo{volume}{3}},
  \bibinfo{pages}{1} (\bibinfo{year}{2017}).

\bibitem[{\citenamefont{Chitambar and Gour}(2019)}]{chitambar2019quantum}
\bibinfo{author}{\bibfnamefont{E.}~\bibnamefont{Chitambar}} \bibnamefont{and}
  \bibinfo{author}{\bibfnamefont{G.}~\bibnamefont{Gour}},
  \bibinfo{journal}{Reviews of Modern Physics} \textbf{\bibinfo{volume}{91}},
  \bibinfo{pages}{025001} (\bibinfo{year}{2019}).

\bibitem[{\citenamefont{Bavaresco et~al.}(2019)\citenamefont{Bavaresco,
  Ara{\'u}jo, Brukner, and Quintino}}]{bavaresco2019semi}
\bibinfo{author}{\bibfnamefont{J.}~\bibnamefont{Bavaresco}},
  \bibinfo{author}{\bibfnamefont{M.}~\bibnamefont{Ara{\'u}jo}},
  \bibinfo{author}{\bibfnamefont{{\v{C}}.}~\bibnamefont{Brukner}},
  \bibnamefont{and} \bibinfo{author}{\bibfnamefont{M.~T.}
  \bibnamefont{Quintino}}, \bibinfo{journal}{Quantum}
  \textbf{\bibinfo{volume}{3}}, \bibinfo{pages}{176} (\bibinfo{year}{2019}).

\bibitem[{\citenamefont{Swingle et~al.}(2016)\citenamefont{Swingle, Bentsen,
  Schleier-Smith, and Hayden}}]{swingle2016measuring}
\bibinfo{author}{\bibfnamefont{B.}~\bibnamefont{Swingle}},
  \bibinfo{author}{\bibfnamefont{G.}~\bibnamefont{Bentsen}},
  \bibinfo{author}{\bibfnamefont{M.}~\bibnamefont{Schleier-Smith}},
  \bibnamefont{and} \bibinfo{author}{\bibfnamefont{P.}~\bibnamefont{Hayden}},
  \bibinfo{journal}{Physical Review A} \textbf{\bibinfo{volume}{94}},
  \bibinfo{pages}{040302(R)} (\bibinfo{year}{2016}).

\bibitem[{\citenamefont{Zhu et~al.}(2016)\citenamefont{Zhu, Hafezi, and
  Grover}}]{zhu2016measurement}
\bibinfo{author}{\bibfnamefont{G.}~\bibnamefont{Zhu}},
  \bibinfo{author}{\bibfnamefont{M.}~\bibnamefont{Hafezi}}, \bibnamefont{and}
  \bibinfo{author}{\bibfnamefont{T.}~\bibnamefont{Grover}},
  \bibinfo{journal}{Physical Review A} \textbf{\bibinfo{volume}{94}},
  \bibinfo{pages}{062329} (\bibinfo{year}{2016}).

\bibitem[{\citenamefont{Nie et~al.}(2020)\citenamefont{Nie, Zhu, Xi, Long, Lin,
  Tian, Qiu, Yang, Dong, Li et~al.}}]{nie2020experimental}
\bibinfo{author}{\bibfnamefont{X.}~\bibnamefont{Nie}},
  \bibinfo{author}{\bibfnamefont{X.}~\bibnamefont{Zhu}},
  \bibinfo{author}{\bibfnamefont{C.}~\bibnamefont{Xi}},
  \bibinfo{author}{\bibfnamefont{X.}~\bibnamefont{Long}},
  \bibinfo{author}{\bibfnamefont{Z.}~\bibnamefont{Lin}},
  \bibinfo{author}{\bibfnamefont{Y.}~\bibnamefont{Tian}},
  \bibinfo{author}{\bibfnamefont{C.}~\bibnamefont{Qiu}},
  \bibinfo{author}{\bibfnamefont{X.}~\bibnamefont{Yang}},
  \bibinfo{author}{\bibfnamefont{Y.}~\bibnamefont{Dong}},
  \bibinfo{author}{\bibfnamefont{J.}~\bibnamefont{Li}}, \bibnamefont{et~al.},
  \bibinfo{journal}{arXiv preprint arXiv:2011.12580}  (\bibinfo{year}{2020}).

\bibitem[{\citenamefont{Ro{\ss}nagel et~al.}(2016)\citenamefont{Ro{\ss}nagel,
  Dawkins, Tolazzi, Abah, Lutz, Schmidt-Kaler, and
  Singer}}]{rossnagel2016single}
\bibinfo{author}{\bibfnamefont{J.}~\bibnamefont{Ro{\ss}nagel}},
  \bibinfo{author}{\bibfnamefont{S.~T.} \bibnamefont{Dawkins}},
  \bibinfo{author}{\bibfnamefont{K.~N.} \bibnamefont{Tolazzi}},
  \bibinfo{author}{\bibfnamefont{O.}~\bibnamefont{Abah}},
  \bibinfo{author}{\bibfnamefont{E.}~\bibnamefont{Lutz}},
  \bibinfo{author}{\bibfnamefont{F.}~\bibnamefont{Schmidt-Kaler}},
  \bibnamefont{and} \bibinfo{author}{\bibfnamefont{K.}~\bibnamefont{Singer}},
  \bibinfo{journal}{Science} \textbf{\bibinfo{volume}{352}},
  \bibinfo{pages}{325} (\bibinfo{year}{2016}).

\bibitem[{\citenamefont{Hu et~al.}(2020)\citenamefont{Hu, Santos, Cui, Huang,
  Soares-Pinto, Sarandy, Li, and Guo}}]{hu2020quantum}
\bibinfo{author}{\bibfnamefont{C.-K.} \bibnamefont{Hu}},
  \bibinfo{author}{\bibfnamefont{A.~C.} \bibnamefont{Santos}},
  \bibinfo{author}{\bibfnamefont{J.-M.} \bibnamefont{Cui}},
  \bibinfo{author}{\bibfnamefont{Y.-F.} \bibnamefont{Huang}},
  \bibinfo{author}{\bibfnamefont{D.~O.} \bibnamefont{Soares-Pinto}},
  \bibinfo{author}{\bibfnamefont{M.~S.} \bibnamefont{Sarandy}},
  \bibinfo{author}{\bibfnamefont{C.-F.} \bibnamefont{Li}}, \bibnamefont{and}
  \bibinfo{author}{\bibfnamefont{G.-C.} \bibnamefont{Guo}},
  \bibinfo{journal}{npj Quantum Information} \textbf{\bibinfo{volume}{6}},
  \bibinfo{pages}{1} (\bibinfo{year}{2020}).

\bibitem[{\citenamefont{Raizen}(2009)}]{raizen2009comprehensive}
\bibinfo{author}{\bibfnamefont{M.~G.} \bibnamefont{Raizen}},
  \bibinfo{journal}{Science} \textbf{\bibinfo{volume}{324}},
  \bibinfo{pages}{1403} (\bibinfo{year}{2009}).

\bibitem[{\citenamefont{Quan et~al.}(2006)\citenamefont{Quan, Wang, Liu, Sun,
  and Nori}}]{quan2006maxwell}
\bibinfo{author}{\bibfnamefont{H.}~\bibnamefont{Quan}},
  \bibinfo{author}{\bibfnamefont{Y.}~\bibnamefont{Wang}},
  \bibinfo{author}{\bibfnamefont{Y.-x.} \bibnamefont{Liu}},
  \bibinfo{author}{\bibfnamefont{C.}~\bibnamefont{Sun}}, \bibnamefont{and}
  \bibinfo{author}{\bibfnamefont{F.}~\bibnamefont{Nori}},
  \bibinfo{journal}{Physical review letters} \textbf{\bibinfo{volume}{97}},
  \bibinfo{pages}{180402} (\bibinfo{year}{2006}).

\bibitem[{\citenamefont{Cottet et~al.}(2017)\citenamefont{Cottet, Jezouin,
  Bretheau, Campagne-Ibarcq, Ficheux, Anders, Auff{\`e}ves, Azouit, Rouchon,
  and Huard}}]{cottet2017observing}
\bibinfo{author}{\bibfnamefont{N.}~\bibnamefont{Cottet}},
  \bibinfo{author}{\bibfnamefont{S.}~\bibnamefont{Jezouin}},
  \bibinfo{author}{\bibfnamefont{L.}~\bibnamefont{Bretheau}},
  \bibinfo{author}{\bibfnamefont{P.}~\bibnamefont{Campagne-Ibarcq}},
  \bibinfo{author}{\bibfnamefont{Q.}~\bibnamefont{Ficheux}},
  \bibinfo{author}{\bibfnamefont{J.}~\bibnamefont{Anders}},
  \bibinfo{author}{\bibfnamefont{A.}~\bibnamefont{Auff{\`e}ves}},
  \bibinfo{author}{\bibfnamefont{R.}~\bibnamefont{Azouit}},
  \bibinfo{author}{\bibfnamefont{P.}~\bibnamefont{Rouchon}}, \bibnamefont{and}
  \bibinfo{author}{\bibfnamefont{B.}~\bibnamefont{Huard}},
  \bibinfo{journal}{Proceedings of the National Academy of Sciences}
  \textbf{\bibinfo{volume}{114}}, \bibinfo{pages}{7561} (\bibinfo{year}{2017}).

\bibitem[{\citenamefont{Chida et~al.}(2017)\citenamefont{Chida, Desai,
  Nishiguchi, and Fujiwara}}]{chida2017power}
\bibinfo{author}{\bibfnamefont{K.}~\bibnamefont{Chida}},
  \bibinfo{author}{\bibfnamefont{S.}~\bibnamefont{Desai}},
  \bibinfo{author}{\bibfnamefont{K.}~\bibnamefont{Nishiguchi}},
  \bibnamefont{and} \bibinfo{author}{\bibfnamefont{A.}~\bibnamefont{Fujiwara}},
  \bibinfo{journal}{Nature communications} \textbf{\bibinfo{volume}{8}},
  \bibinfo{pages}{1} (\bibinfo{year}{2017}).

\bibitem[{\citenamefont{Camati et~al.}(2016)\citenamefont{Camati, Peterson,
  Batalhao, Micadei, Souza, Sarthour, Oliveira, and
  Serra}}]{camati2016experimental}
\bibinfo{author}{\bibfnamefont{P.~A.} \bibnamefont{Camati}},
  \bibinfo{author}{\bibfnamefont{J.~P.~S.} \bibnamefont{Peterson}},
  \bibinfo{author}{\bibfnamefont{T.~B.} \bibnamefont{Batalhao}},
  \bibinfo{author}{\bibfnamefont{K.}~\bibnamefont{Micadei}},
  \bibinfo{author}{\bibfnamefont{A.~M.} \bibnamefont{Souza}},
  \bibinfo{author}{\bibfnamefont{R.~S.} \bibnamefont{Sarthour}},
  \bibinfo{author}{\bibfnamefont{I.~S.} \bibnamefont{Oliveira}},
  \bibnamefont{and} \bibinfo{author}{\bibfnamefont{R.~M.} \bibnamefont{Serra}},
  \bibinfo{journal}{Physical Review Letters} \textbf{\bibinfo{volume}{117}},
  \bibinfo{pages}{240502} (\bibinfo{year}{2016}).

\bibitem[{\citenamefont{Ringbauer et~al.}(2018)\citenamefont{Ringbauer,
  Bromley, Cianciaruso, Lami, Lau, Adesso, White, Fedrizzi, and
  Piani}}]{ringbauer2018certification}
\bibinfo{author}{\bibfnamefont{M.}~\bibnamefont{Ringbauer}},
  \bibinfo{author}{\bibfnamefont{T.~R.} \bibnamefont{Bromley}},
  \bibinfo{author}{\bibfnamefont{M.}~\bibnamefont{Cianciaruso}},
  \bibinfo{author}{\bibfnamefont{L.}~\bibnamefont{Lami}},
  \bibinfo{author}{\bibfnamefont{W.~S.} \bibnamefont{Lau}},
  \bibinfo{author}{\bibfnamefont{G.}~\bibnamefont{Adesso}},
  \bibinfo{author}{\bibfnamefont{A.~G.} \bibnamefont{White}},
  \bibinfo{author}{\bibfnamefont{A.}~\bibnamefont{Fedrizzi}}, \bibnamefont{and}
  \bibinfo{author}{\bibfnamefont{M.}~\bibnamefont{Piani}},
  \bibinfo{journal}{Physical Review X} \textbf{\bibinfo{volume}{8}},
  \bibinfo{pages}{041007} (\bibinfo{year}{2018}).

\bibitem[{\citenamefont{Ramelow et~al.}(2009)\citenamefont{Ramelow,
  Ratschbacher, Fedrizzi, Langford, and Zeilinger}}]{ramelow2009discrete}
\bibinfo{author}{\bibfnamefont{S.}~\bibnamefont{Ramelow}},
  \bibinfo{author}{\bibfnamefont{L.}~\bibnamefont{Ratschbacher}},
  \bibinfo{author}{\bibfnamefont{A.}~\bibnamefont{Fedrizzi}},
  \bibinfo{author}{\bibfnamefont{N.}~\bibnamefont{Langford}}, \bibnamefont{and}
  \bibinfo{author}{\bibfnamefont{A.}~\bibnamefont{Zeilinger}},
  \bibinfo{journal}{Physical review letters} \textbf{\bibinfo{volume}{103}},
  \bibinfo{pages}{253601} (\bibinfo{year}{2009}).

\bibitem[{\citenamefont{Reimer et~al.}(2016)\citenamefont{Reimer, Kues,
  Roztocki, Wetzel, Grazioso, Little, Chu, Johnston, Bromberg, Caspani
  et~al.}}]{reimer2016generation}
\bibinfo{author}{\bibfnamefont{C.}~\bibnamefont{Reimer}},
  \bibinfo{author}{\bibfnamefont{M.}~\bibnamefont{Kues}},
  \bibinfo{author}{\bibfnamefont{P.}~\bibnamefont{Roztocki}},
  \bibinfo{author}{\bibfnamefont{B.}~\bibnamefont{Wetzel}},
  \bibinfo{author}{\bibfnamefont{F.}~\bibnamefont{Grazioso}},
  \bibinfo{author}{\bibfnamefont{B.~E.} \bibnamefont{Little}},
  \bibinfo{author}{\bibfnamefont{S.~T.} \bibnamefont{Chu}},
  \bibinfo{author}{\bibfnamefont{T.}~\bibnamefont{Johnston}},
  \bibinfo{author}{\bibfnamefont{Y.}~\bibnamefont{Bromberg}},
  \bibinfo{author}{\bibfnamefont{L.}~\bibnamefont{Caspani}},
  \bibnamefont{et~al.}, \bibinfo{journal}{Science}
  \textbf{\bibinfo{volume}{351}}, \bibinfo{pages}{1176} (\bibinfo{year}{2016}).

\bibitem[{\citenamefont{Kues et~al.}(2017)\citenamefont{Kues, Reimer, Roztocki,
  Cort{\'e}s, Sciara, Wetzel, Zhang, Cino, Chu, Little et~al.}}]{kues2017chip}
\bibinfo{author}{\bibfnamefont{M.}~\bibnamefont{Kues}},
  \bibinfo{author}{\bibfnamefont{C.}~\bibnamefont{Reimer}},
  \bibinfo{author}{\bibfnamefont{P.}~\bibnamefont{Roztocki}},
  \bibinfo{author}{\bibfnamefont{L.~R.} \bibnamefont{Cort{\'e}s}},
  \bibinfo{author}{\bibfnamefont{S.}~\bibnamefont{Sciara}},
  \bibinfo{author}{\bibfnamefont{B.}~\bibnamefont{Wetzel}},
  \bibinfo{author}{\bibfnamefont{Y.}~\bibnamefont{Zhang}},
  \bibinfo{author}{\bibfnamefont{A.}~\bibnamefont{Cino}},
  \bibinfo{author}{\bibfnamefont{S.~T.} \bibnamefont{Chu}},
  \bibinfo{author}{\bibfnamefont{B.~E.} \bibnamefont{Little}},
  \bibnamefont{et~al.}, \bibinfo{journal}{Nature}
  \textbf{\bibinfo{volume}{546}}, \bibinfo{pages}{622} (\bibinfo{year}{2017}).

\bibitem[{\citenamefont{Lu et~al.}(2020)\citenamefont{Lu, Simmerman, Lougovski,
  Weiner, and Lukens}}]{lu2020fully}
\bibinfo{author}{\bibfnamefont{H.-H.} \bibnamefont{Lu}},
  \bibinfo{author}{\bibfnamefont{E.~M.} \bibnamefont{Simmerman}},
  \bibinfo{author}{\bibfnamefont{P.}~\bibnamefont{Lougovski}},
  \bibinfo{author}{\bibfnamefont{A.~M.} \bibnamefont{Weiner}},
  \bibnamefont{and} \bibinfo{author}{\bibfnamefont{J.~M.}
  \bibnamefont{Lukens}}, \bibinfo{journal}{Physical Review Letters}
  \textbf{\bibinfo{volume}{125}}, \bibinfo{pages}{120503}
  (\bibinfo{year}{2020}).

\bibitem[{\citenamefont{Abbott et~al.}(2020)\citenamefont{Abbott, Wechs,
  Horsman, Mhalla, and Branciard}}]{abbott2020communication}
\bibinfo{author}{\bibfnamefont{A.~A.} \bibnamefont{Abbott}},
  \bibinfo{author}{\bibfnamefont{J.}~\bibnamefont{Wechs}},
  \bibinfo{author}{\bibfnamefont{D.}~\bibnamefont{Horsman}},
  \bibinfo{author}{\bibfnamefont{M.}~\bibnamefont{Mhalla}}, \bibnamefont{and}
  \bibinfo{author}{\bibfnamefont{C.}~\bibnamefont{Branciard}},
  \bibinfo{journal}{Quantum} \textbf{\bibinfo{volume}{4}}, \bibinfo{pages}{333}
  (\bibinfo{year}{2020}).

\bibitem[{\citenamefont{Gu{\'e}rin et~al.}(2019)\citenamefont{Gu{\'e}rin,
  Rubino, and Brukner}}]{guerin2019communication}
\bibinfo{author}{\bibfnamefont{P.~A.} \bibnamefont{Gu{\'e}rin}},
  \bibinfo{author}{\bibfnamefont{G.}~\bibnamefont{Rubino}}, \bibnamefont{and}
  \bibinfo{author}{\bibfnamefont{{\v{C}}.}~\bibnamefont{Brukner}},
  \bibinfo{journal}{Physical Review A} \textbf{\bibinfo{volume}{99}},
  \bibinfo{pages}{062317} (\bibinfo{year}{2019}).

\bibitem[{\citenamefont{Chiribella and
  Kristj{\'a}nsson}(2019)}]{chiribella2019quantum}
\bibinfo{author}{\bibfnamefont{G.}~\bibnamefont{Chiribella}} \bibnamefont{and}
  \bibinfo{author}{\bibfnamefont{H.}~\bibnamefont{Kristj{\'a}nsson}},
  \bibinfo{journal}{Proceedings of the Royal Society A}
  \textbf{\bibinfo{volume}{475}}, \bibinfo{pages}{20180903}
  (\bibinfo{year}{2019}).

\bibitem[{\citenamefont{Rubino et~al.}(2021)\citenamefont{Rubino, Rozema,
  Ebler, Kristj{\'a}nsson, Salek, Gu{\'e}rin, Abbott, Branciard, Brukner,
  Chiribella et~al.}}]{rubino2021experimental}
\bibinfo{author}{\bibfnamefont{G.}~\bibnamefont{Rubino}},
  \bibinfo{author}{\bibfnamefont{L.~A.} \bibnamefont{Rozema}},
  \bibinfo{author}{\bibfnamefont{D.}~\bibnamefont{Ebler}},
  \bibinfo{author}{\bibfnamefont{H.}~\bibnamefont{Kristj{\'a}nsson}},
  \bibinfo{author}{\bibfnamefont{S.}~\bibnamefont{Salek}},
  \bibinfo{author}{\bibfnamefont{P.~A.} \bibnamefont{Gu{\'e}rin}},
  \bibinfo{author}{\bibfnamefont{A.~A.} \bibnamefont{Abbott}},
  \bibinfo{author}{\bibfnamefont{C.}~\bibnamefont{Branciard}},
  \bibinfo{author}{\bibfnamefont{{\v{C}}.}~\bibnamefont{Brukner}},
  \bibinfo{author}{\bibfnamefont{G.}~\bibnamefont{Chiribella}},
  \bibnamefont{et~al.}, \bibinfo{journal}{Physical Review Research}
  \textbf{\bibinfo{volume}{3}}, \bibinfo{pages}{013093} (\bibinfo{year}{2021}).

\bibitem[{\citenamefont{Gisin et~al.}(2005)\citenamefont{Gisin, Linden, Massar,
  and Popescu}}]{gisin2005error}
\bibinfo{author}{\bibfnamefont{N.}~\bibnamefont{Gisin}},
  \bibinfo{author}{\bibfnamefont{N.}~\bibnamefont{Linden}},
  \bibinfo{author}{\bibfnamefont{S.}~\bibnamefont{Massar}}, \bibnamefont{and}
  \bibinfo{author}{\bibfnamefont{S.}~\bibnamefont{Popescu}},
  \bibinfo{journal}{Physical Review A} \textbf{\bibinfo{volume}{72}},
  \bibinfo{pages}{012338} (\bibinfo{year}{2005}).

\bibitem[{\citenamefont{Ara{\'u}jo
  et~al.}(2014{\natexlab{b}})\citenamefont{Ara{\'u}jo, Feix, Costa, and
  Brukner}}]{araujo2014quantum}
\bibinfo{author}{\bibfnamefont{M.}~\bibnamefont{Ara{\'u}jo}},
  \bibinfo{author}{\bibfnamefont{A.}~\bibnamefont{Feix}},
  \bibinfo{author}{\bibfnamefont{F.}~\bibnamefont{Costa}}, \bibnamefont{and}
  \bibinfo{author}{\bibfnamefont{{\v{C}}.}~\bibnamefont{Brukner}},
  \bibinfo{journal}{New Journal of Physics} \textbf{\bibinfo{volume}{16}},
  \bibinfo{pages}{093026} (\bibinfo{year}{2014}{\natexlab{b}}).

\bibitem[{\citenamefont{Wood et~al.}(2021)\citenamefont{Wood, Verma, Costa, and
  Zych}}]{wood2021operational}
\bibinfo{author}{\bibfnamefont{C.~E.} \bibnamefont{Wood}},
  \bibinfo{author}{\bibfnamefont{H.}~\bibnamefont{Verma}},
  \bibinfo{author}{\bibfnamefont{F.}~\bibnamefont{Costa}}, \bibnamefont{and}
  \bibinfo{author}{\bibfnamefont{M.}~\bibnamefont{Zych}},
  \bibinfo{journal}{arXiv preprint arXiv:2112.07860}  (\bibinfo{year}{2021}).

\bibitem[{\citenamefont{Nie et~al.}(2022)\citenamefont{Nie, Feng, Longden, and
  Vedral}}]{nie2022quantum}
\bibinfo{author}{\bibfnamefont{H.}~\bibnamefont{Nie}},
  \bibinfo{author}{\bibfnamefont{T.}~\bibnamefont{Feng}},
  \bibinfo{author}{\bibfnamefont{S.}~\bibnamefont{Longden}}, \bibnamefont{and}
  \bibinfo{author}{\bibfnamefont{V.}~\bibnamefont{Vedral}},
  \bibinfo{journal}{arXiv preprint arXiv:2201.06954}  (\bibinfo{year}{2022}).

\bibitem[{\citenamefont{Landauer et~al.}(1991)}]{landauer1991information}
\bibinfo{author}{\bibfnamefont{R.}~\bibnamefont{Landauer}}
  \bibnamefont{et~al.}, \bibinfo{journal}{Physics Today}
  \textbf{\bibinfo{volume}{44}}, \bibinfo{pages}{23} (\bibinfo{year}{1991}).

\bibitem[{\citenamefont{Callen}(1985)}]{callen1985thermodynamics}
\bibinfo{author}{\bibfnamefont{H.}~\bibnamefont{Callen}}, \bibinfo{journal}{New
  York}  (\bibinfo{year}{1985}).

\bibitem[{\citenamefont{Sagawa and Ueda}(2010)}]{sagawa2010generalized}
\bibinfo{author}{\bibfnamefont{T.}~\bibnamefont{Sagawa}} \bibnamefont{and}
  \bibinfo{author}{\bibfnamefont{M.}~\bibnamefont{Ueda}},
  \bibinfo{journal}{Physical review letters} \textbf{\bibinfo{volume}{104}},
  \bibinfo{pages}{090602} (\bibinfo{year}{2010}).

\bibitem[{\citenamefont{Sagawa and Ueda}(2008)}]{sagawa2008second}
\bibinfo{author}{\bibfnamefont{T.}~\bibnamefont{Sagawa}} \bibnamefont{and}
  \bibinfo{author}{\bibfnamefont{M.}~\bibnamefont{Ueda}},
  \bibinfo{journal}{Physical review letters} \textbf{\bibinfo{volume}{100}},
  \bibinfo{pages}{080403} (\bibinfo{year}{2008}).

\bibitem[{\citenamefont{Elouard and Jordan}(2018)}]{elouard2018efficient}
\bibinfo{author}{\bibfnamefont{C.}~\bibnamefont{Elouard}} \bibnamefont{and}
  \bibinfo{author}{\bibfnamefont{A.~N.} \bibnamefont{Jordan}},
  \bibinfo{journal}{Physical review letters} \textbf{\bibinfo{volume}{120}},
  \bibinfo{pages}{260601} (\bibinfo{year}{2018}).

\bibitem[{\citenamefont{Elouard
  et~al.}(2017{\natexlab{b}})\citenamefont{Elouard, Herrera-Mart{\'\i}, Clusel,
  and Auffeves}}]{elouard2017role}
\bibinfo{author}{\bibfnamefont{C.}~\bibnamefont{Elouard}},
  \bibinfo{author}{\bibfnamefont{D.~A.} \bibnamefont{Herrera-Mart{\'\i}}},
  \bibinfo{author}{\bibfnamefont{M.}~\bibnamefont{Clusel}}, \bibnamefont{and}
  \bibinfo{author}{\bibfnamefont{A.}~\bibnamefont{Auffeves}},
  \bibinfo{journal}{npj Quantum Information} \textbf{\bibinfo{volume}{3}},
  \bibinfo{pages}{1} (\bibinfo{year}{2017}{\natexlab{b}}).

\bibitem[{\citenamefont{Koski et~al.}(2014{\natexlab{a}})\citenamefont{Koski,
  Maisi, Sagawa, and Pekola}}]{koski2014experimental}
\bibinfo{author}{\bibfnamefont{J.~V.} \bibnamefont{Koski}},
  \bibinfo{author}{\bibfnamefont{V.~F.} \bibnamefont{Maisi}},
  \bibinfo{author}{\bibfnamefont{T.}~\bibnamefont{Sagawa}}, \bibnamefont{and}
  \bibinfo{author}{\bibfnamefont{J.~P.} \bibnamefont{Pekola}},
  \bibinfo{journal}{Physical review letters} \textbf{\bibinfo{volume}{113}},
  \bibinfo{pages}{030601} (\bibinfo{year}{2014}{\natexlab{a}}).

\bibitem[{\citenamefont{Koski et~al.}(2014{\natexlab{b}})\citenamefont{Koski,
  Maisi, Pekola, and Averin}}]{koski2014experimentalreal}
\bibinfo{author}{\bibfnamefont{J.~V.} \bibnamefont{Koski}},
  \bibinfo{author}{\bibfnamefont{V.~F.} \bibnamefont{Maisi}},
  \bibinfo{author}{\bibfnamefont{J.~P.} \bibnamefont{Pekola}},
  \bibnamefont{and} \bibinfo{author}{\bibfnamefont{D.~V.}
  \bibnamefont{Averin}}, \bibinfo{journal}{Proceedings of the National Academy
  of Sciences} \textbf{\bibinfo{volume}{111}}, \bibinfo{pages}{13786}
  (\bibinfo{year}{2014}{\natexlab{b}}).

\end{thebibliography}

\end{document}